\documentclass[twocolumn]{aastex61}

\usepackage[normalem]{ulem}
\usepackage{color}
\usepackage{natbib,aas_macros}

\newcommand{\og}{g_{0}}
\newcommand{\oi}{i_{0}}
\newcommand{\ogi}{(g-i)_{0}}
\newcommand{\gitrgb}{(g-i)_{\rm{TRGB}}}

\newcommand{\mv}{\rm{M_{V,0}}}
\newcommand{\vi}{\rm{(V-I)_{0}}}
\newcommand{\mg}{\rm{M_{g}}}
\newcommand{\mm}{(m-M)_{0}}
\newcommand{\gmh}{\rm{[M/H]}}
\newcommand{\feh}{\rm{[Fe/H]}}
\newcommand{\rr}{\rm{R_{25}}}
\newcommand{\rc}{\rm{R_{c}}}
\newcommand{\rt}{\rm{R_{t}}}
\newcommand{\re}{\rm{R_{exp}}}
\newcommand{\rh}{\rm{R_{h}}}
\newcommand{\sqdeg}{\rm{deg^2}}
\newcommand{\msun}{\rm{M}_{\odot}}

\shorttitle{Dwarf galaxies of the M81 Group}
\shortauthors{Okamoto et al.}

\begin{document}

\title{Stellar population and structural properties of Dwarf galaxies and young stellar systems in the M81 group\footnote{Based on data collected at Subaru Telescope, which is operated by the National Astronomical Observatory of Japan.}}

\author{Sakurako Okamoto}
\email{sakurako.okamoto@nao.ac.jp}
\affil{Subaru Telescope, National Astronomical Observatory of Japan, 650 North A'ohoku Place, Hilo, HI 96720, U.S.A.}
\affil{National Astronomical Observatory of Japan, Osawa 2-21-1, Mitaka, Tokyo, 181-8588, JAPAN}
\affil{The Graduate University for Advanced Studies, Osawa 2-21-1, Mitaka, Tokyo 181-8588, JAPAN}
\author{Nobuo Arimoto}
\affil{Subaru Telescope, National Astronomical Observatory of Japan, 650 North A'ohoku Place, Hilo, HI 96720, U.S.A.}
\affil{Astronomy Program, Department of Physics and Astronomy, Seoul National University, 599 Gwanak-ro, Gwanak-gu, Seoul, 151-742, Korea}
\author{Annette M.N. Ferguson}
\affil{Institute for Astronomy, University of Edinburgh, Royal Observatory, Blackford Hill, Edinburgh, EH9 3HJ U.K.}
\author{Mike J. Irwin}
\affil{Institute of Astronomy, University of Cambridge, Madingley Road, Cambridge CB3 0HA, U.K.}
\author{Edouard J. Bernard}
\affil{Observatoire de la C\^ote d'Azur, 96 Boulevard de l'Observatoire, 06300 Nice, France}
\author{Yousuke Utsumi}
\affil{Kavli Institute for Particle Astrophysics and Cosmology, Stanford University, Menlo Park CA 94025, U.S.A.}
\affil{Hiroshima Astrophysical Science Center, Hiroshima University, Kagamiyama 1-3-1, Higashi-Hiroshima, Hiroshima 739-8526, Japan}

\begin{abstract}

We use Hyper Suprime-Cam on the Subaru Telescope to investigate the structural and photometric properties of early-type dwarf galaxies and young stellar systems at the center of the M81 Group.  We have mapped resolved stars to $\sim2$ magnitudes below the tip of the red giant branch over almost 6.5 square degrees,  corresponding to a projected area of $160\times160 \rm{kpc}$ at the distance of M81.  The resulting stellar catalogue enables a homogeneous analysis of the member galaxies with unprecedented sensitivity to low surface brightness emission.  The radial profiles of the dwarf galaxies are well-described by Sersic and King profiles, and show no obvious signatures of tidal disruption.  The measured radii for most of these systems are larger than the existing  
literature values and we find the total luminosity of IKN ($\mv=-14.29$) to be almost 3 magnitudes brighter than previously-thought.  
We identify new dwarf satellite candidates, d1006+69 and d1009+68, which we estimate to lie at a distance of $4.3\pm0.2$ Mpc and $3.5\pm0.5$ Mpc. With $\mv=-8.91\pm0.40$ and $\gmh=-1.83\pm0.28$, d1006+69 is one of the faintest and most metal-poor dwarf satellites currently-known in the M81 Group.  The luminosity functions of young stellar systems in the outlying tidal HI debris imply continuous star formation in the recent past and the existence of populations as
young as 30 Myr old.  We find no evidence for old RGB stars coincident with the young MS/cHeB stars which define these objects, supporting the idea that they are genuinely new stellar systems resulting from triggered star formation in gaseous tidal debris.
\end{abstract}

\keywords{galaxies: groups: individual (M81) --- galaxies: individual (IKN, KDG61, KDG064, BK5N) --- galaxies: stellar content --- galaxies: structure}

\section{Introduction} \label{sec: intro}

Lying at D$\sim3.6$~Mpc, the M81 Group is one of the nearest galaxy groups and its proximity and resemblance to the Local Group have fuelled much research over several decades \citep[e.g.][]{1968ApJ...151..825T, 1971ApJS...22..445S, 1981MNRAS.195..327A, 1994Natur.372..530Y, 2002A&A...383..125K, 2009AJ....138.1469B, 2013AJ....146..126C}.  It contains more than 40 member galaxies, including the large Sb spiral galaxy M81, the peculiar galaxies M82 and NGC3077, 9 late-type galaxies, at least 20 low-luminosity early-type dwarfs and a variety of stellar debris features, some of which are tidal dwarf galaxy candidates.  Most of the member galaxies cluster around either M81 or NGC2403, the low mass spiral galaxy that lies $\sim 14$ degrees to the southwest of M81 and $\sim 1$ Mpc in the foreground. These two sub-groups are moving toward each other \citep{2002A&A...383..125K, 2013AJ....146..126C}.  While the binary sub-group structure and the total number of member galaxies are similar to those of the Local Group, the recent strong gravitational interactions amongst M81, M82 and NGC3077 provide a quite different environment to that found locally.  

These interactions were first revealed through studies of the 21cm emission line of neutral hydrogen (HI) which showed an extremely disturbed HI gas distribution with tidal bridges connecting the three galaxies \citep[e.g.][]{1975ApJ...195...23G, 1981MNRAS.195..327A, 1994Natur.372..530Y, 2008AJ....135.1983C, 2018ApJ...865...26D}.  Numerical simulations by \citet{1999IAUS..186...81Y} successfully reproduced the HI tidal features if the closest encounters of M82 and NGC 3077 to M81 had taken place 220 and 280 Myr ago, respectively.  Recently, \citet{2017MNRAS.467..273O} have re-examined the dynamical history of the M81 Group through conducting a statistical study of the possible orbits of three galaxies, taking into account dynamical friction.  They argue that the most likely explanation for why the galaxies have not already coalesced into a single system is that they were previously unbound and have only recently encountered each other within the past 500 Myr.  On the other hand, if the triplet has been gravitationally bound for a significant time, they
predict that the three systems will likely merge within the next 1-2 Gyr.

Some of the prominent outlying HI tidal features around M81 are well-known to have associated young
stellar counterparts, for e.g.  the Garland, Holmberg IX and Arp's Loop. Deep photometry from the Hubble Space Telescope (HST), and from the ground, has been used to argue that recent interactions have induced star formation in these regions and that they are some of the closest examples of candidate tidal dwarf galaxies \citep{2002A&A...396..473M, 2008AJ....135..548D, 2008ApJ...676L.113S}.  

A number of new galaxies of the M81 Group have recently been discovered via their resolved stellar populations.  \citet{2009AJ....137.3009C} found 22 candidates within a $65~\sqdeg$ area around M81 using MegaCam on Canada-France-Hawaii Telescope, of which 14 were confirmed as dwarf galaxy members with deep HST photometry  \citep{2013AJ....146..126C}.  \citet{2017ApJ...843L...6S} discovered another system in the vicinity of BK5N using images taken with Hyper Suprime-Cam (HSC) on the Subaru Telescope.  To date, more than 20 faint dwarf galaxies are known as members of the M81 Group, but
detailed studies of these systems are still rather limited. 

We are conducting a deep contiguous photometric survey of resolved stars in the M81 Group using HSC on the Subaru Telescope. When complete, our survey will cover approximately 7.5 sq. degrees and will map stars out to a radius of 120 kpc from the M81 galaxy center.  Figure \ref{fig: pointing} displays the sky coverage of our survey, with colour-coding indicating the current status. Specifically, we have four HSC pointings completely reduced (blue solid circles), a further pointing has been observed and is currently under analysis (green dotted circle), while two remaining fields are yet to be observed (dashed magenta circle). Early highlights from this survey include the discovery of very extended (and rather perturbed) stellar halos around M81, M82 and NGC3077 as well as an enormous tidal stream which links the three systems    \citep{2015ApJ...809L...1O}.

In the present paper, we conduct a homogeneous analysis of the dwarf galaxies and young stellar systems that fall within the four HSC pointings that we have analysed so far. In section \ref{sec: obs}, the details of observations and the data reduction are described.  The photometric and structural properties of the early-type dwarf galaxies and two new galaxy candidates, d1006+69 and d1009+68, are examined, based on their resolved stars in Section \ref{sec: old dwarfs} and Section \ref{new dwarf}, respectively.  In Section \ref{sec: young dwarfs}, we investigate the properties of young stellar systems within the current survey footprint.  The tidal effects on the early-type dwarfs, the comparison with dwarf galaxies in the nearby universe, and the old stellar constituents at the same location as the young systems are discussed in Section \ref{sec: discussion}.  Finally, Section \ref{sec: summary} summarises our findings.

\section{Observations and data reduction} \label{sec: obs}

\begin{figure}
\begin{center}
 \includegraphics[width=240pt]{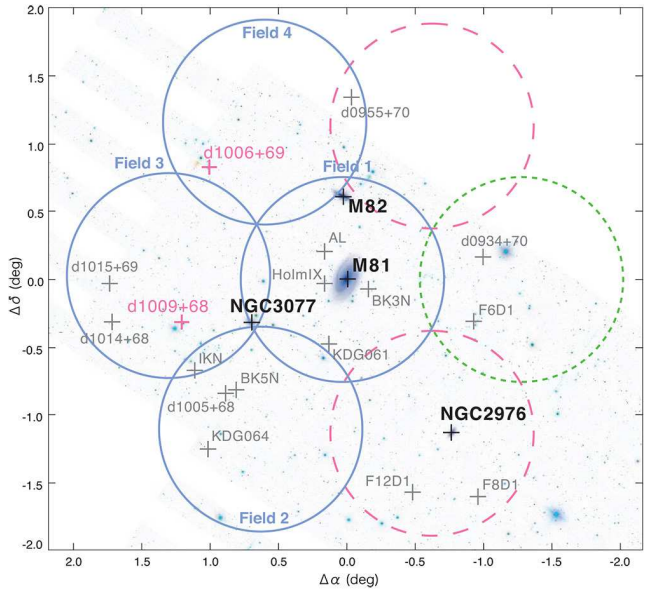}
 \caption{The target pointings of the HSC survey overlaid on an image of an SDSS image of M81 \citep{2000AJ....120.1579Y}.  The circles represent the field-of-view of HSC.  The fields used in this study are marked with blue circles, while fields for which we are still processing or obtaining data are shown as the dotted green and dashed magenta circles, respectively.  The centers of M81, M82, NGC3077 and NGC2976 are shown as the black crosses. } The known member galaxies of the M81 Group and two new dwarf candidates d1006+69 and d1009+68 are displayed as gray and magenta crosses, respectively. 
 \label{fig: pointing}
\end{center}
\end{figure}

The images we use for this study were taken in the $g$- and $i$-band filters under photometric conditions on January 21 and 22, 2015. The seeing ranged from 0.6\arcsec to 0.9\arcsec.  We took several dithered sub-exposures with small overlapping regions to cover the inter-chip gaps and to ensure photometric consistency. The details of the observations are listed in Table \ref{tbl: obs log}. In total, a projected area of 25,200 $\rm{kpc}^2$ at the distance of M81 was mapped, which encompasses a number of known member galaxies as displayed in Figure \ref{fig: pointing}.  In this article, we adopt a distance modulus for M81 and associated systems of $\mm=27.79$ \citep{2011ApJS..195...18R};

\begin{deluxetable}{rccccc}
\tabletypesize{\scriptsize}
\tablecaption{The Observations\label{tbl: obs log}}
\tablehead{
\colhead{Field}     & \colhead{RA(J2000)}    &  \colhead{Dec(J2000)} &
\colhead{Filter}    & \colhead{Exposure}     &  \colhead{seeing} }
\startdata
Field 1 & $09^{h}55^{m}33^{s}.20$ & $+69\arcdeg03\arcmin55\arcsec.00$ & $g$  & 22$\times$200s & 0\farcs87 \\
       &             &              & $i$ &  34$\times$210s & 0\farcs80 \\
\tableline
Field 2 & $10^{h}02^{m}33^{s}.01$ & $+67\arcdeg58\arcmin57\arcsec.90$ & $g$  & 20$\times$200s & 0\farcs74 \\
       &             &              & $i$ &  30$\times$210s & 0\farcs73 \\
\tableline
Field 3 & $10^{h}09^{m}32^{s}.82$ & $+69\arcdeg03\arcmin55\arcsec.00$ & $g$  & 25$\times$200s & 0\farcs94 \\
       &             &              & $i$ &  27$\times$210s & 0\farcs85 \\
\tableline
Field 4 & $10^{h}02^{m}33^{s}.01$ & $+70\arcdeg08\arcmin52\arcsec.12$ & $g$  & 20$\times$200s & 0\farcs70 \\
       &             &              & $i$ &  30$\times$210s & 0\farcs65 \\
\enddata
\end{deluxetable}

The HSC imager consists of a mosaic of 104 CCDs and provides a field-of-view of 1.76 $\sqdeg$ with a pixel scale of 0.17\arcsec \citep{2012SPIE.8446E..0ZM}. Raw data for which the seeing was $<1\arcsec$ or better were processed in the usual manner using the HSC pipeline version 4.0.0 \citet{2018PASJ...70S...5B}, which is based on the software suite being developed for the Large Synoptic Survey Telescope (LSST) project \citep{2008arXiv0805.2366I, 2010SPIE.7740E..15A}.  The pipeline accomplishes the bias correction, flat fielding, mosaicking, stacking and the calibrations.  Astrometric and photometric calibrations were done using a stellar catalogue from the Panoramic Survey Telescope and Rapid Response System (Pan-STARRS) 1 survey \citep{2012ApJ...756..158S,2012ApJ...750...99T,2013ApJS..205...20M} with the
final photometry on the HSC filter system and in AB magnitudes. 

\begin{figure}
\begin{center}
 \includegraphics[width=200pt]{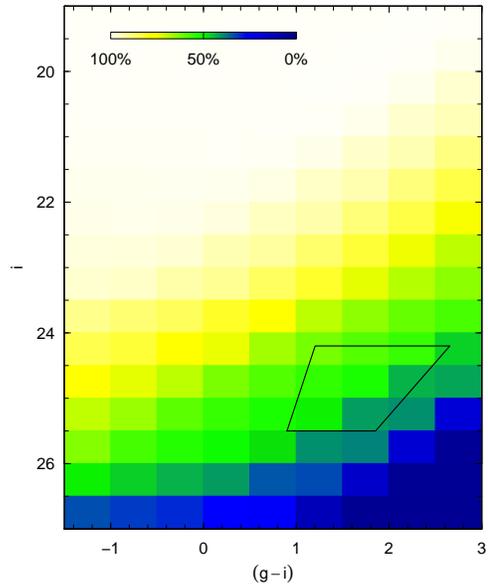}
 \caption{The completeness of point sources in a region of Field 2, shown as a function of $(g-i)$ and $i$-band magnitude. The solid line delineates the selection criterion of RGB stars at the distance of M81. The bin size is 0.5 mag on a side. }
 \label{fig: completeness}
\end{center}
\end{figure}

\begin{figure*}
\begin{center}
 \includegraphics[width=530pt]{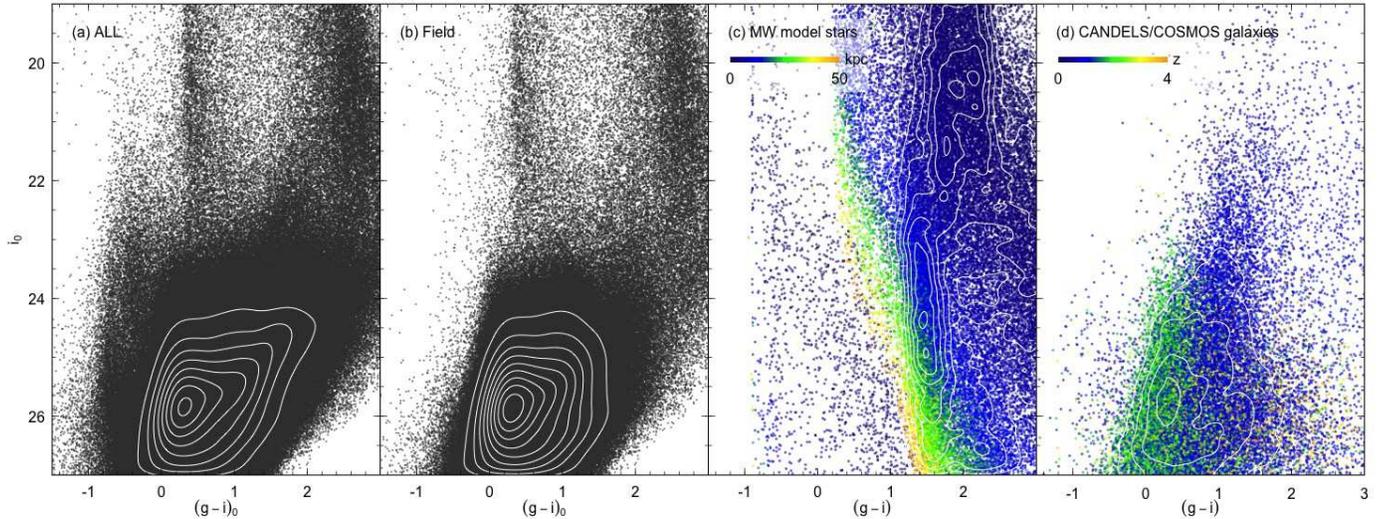}
 \vspace{5pt}
 \caption{The comparison of point sources in our HSC survey with simulated Milky Way populations and distant galaxies in colour-magnitude space. (a) The CMD of point sources in HSC footprint.  (b) The CMD of point sources located outside of $4\times\rm{R}_{25}$ radii of M81 and M82, and outside the tidal radii of the dwarf galaxies.  (c) The CMD of simulated Galactic foreground stars toward the direction of M81, generated with TRILEGAL.  The colour of each star represents its Galactocentric distance. (d) The CMD of galaxies in the CANDELS/COSMOS catalogue \citep{2017ApJS..228....7N}. The colour of each galaxy indicates the photometric redshift.}
 \label{fig: cmds of contaminations}
\end{center}
\end{figure*}

We used the IRAF implementation of DAOPHOT \citep{1987PASP...99..191S} to obtain point spread function (PSF) photometry of sources across the survey area.  Photometry was carried out on 4200 pixel $\times$ 4200 pixel patches of the stacked images of Field 1 to 4  in each filter.   In order to separate point sources from the extended objects and noise-like detections, we used the DAOPHOT parameters $\chi^2$ and SHARP in the same manner as in our previous studies \citep[e.g.][]{2012ApJ...744...96O}.  Specifically, we retained objects whose $chi$ and SHARP values lie within $3\sigma$ of the mean values of the artificial stars (described below) of the same magnitude.  The positions of the detected stellar objects in each patch and in each passband were cross-matched within 0.\arcsec4 to make a composite band-merged catalogue,
containing over 1.3 million objects.  Note that the data previously-presented in \cite{2015ApJ...809L...1O} were re-processed and re-calibrated to ensure consistency in the final catalogue, which now covers about twice the area as the previous one. 

To evaluate the photometric uncertainty and the source detection limit, artificial star tests were performed on several patches in Fields 1 to 4 using the ADDSTAR task in DAOPHOT.  Figure \ref{fig: completeness} demonstrates the resulting photometric completeness as a function of magnitude and colour for a typical field.  The point source catalogue is at least 50$\%$ complete to 26 mag in both passbands and at the faint limit of our RGB selection box, except for the inner crowded regions of galaxies.  We de-redden each source individually according to its location using the Galactic extinction map of \citet{1998ApJ...500..525S} calibrated by \citet{2012ApJ...756..158S}, assuming a \citet{1999PASP..111...63F} reddening law with $\rm{R_{V}}=3.1$. 

Figure \ref{fig: cmds of contaminations}a shows all objects in the final point source catalogue, which includes genuine stars in the M81 Group galaxies, foreground Milky Way stars and unresolved background galaxies that are mis-classified as point sources by our star/galaxy separation.  Although these various populations overlap to some extent in the colour-magnitude diagram (CMD), the well-populated red giant branch (RGB), and its tip at the distance of M81, is clearly delineated  by the extended contour at $\oi\sim24.5$ and at $\ogi>1$, which is not seen in Figure \ref{fig: cmds of contaminations}b.  

To examine the foreground/background contaminants,  we define a ``Field" area to be the regions lying beyond $4\times\rm{R}_{25}$ radii of M81 ($\rm{r}>55.\arcmin2$) and M82 ($\rm{r}>22.\arcmin4$), and beyond the tidal radii of the dwarf galaxies, as estimated in Section \ref{sec: old dwarfs} and in the forthcoming paper for NGC3077 (Okamoto et al. in prep).  We compare objects in the Field area, which covers about  $4.5~\sqdeg$,  with a simulated Milky Way population and with a deep extragalactic field in Figure \ref{fig: cmds of contaminations}b-\ref{fig: cmds of contaminations}d.  The ``Field" population CMD is shown in Figure \ref{fig: cmds of contaminations}b where the dominant populations are seen to be (i) the vertical distribution at $\ogi \sim 0.3$ and $\oi<23$, (ii) the vertical distribution at $\ogi>2.0$ and $\oi<24$, and (iii) the vast number of objects at $\oi>23$ that become bluer at fainter magnitudes.  

We generated a mock catalogue of Galactic stars in a $4.5~\sqdeg$ area in the direction of M81 using TRILEGAL 1.6\footnote{http://stev.oapd.inaf.it/cgi-bin/trilegal} \citep{2012ASSP...26..165G} and show the results in Figure \ref{fig: cmds of contaminations}c with colour-coding according to Galactocentric distance.  Compared with the distribution of Galactic stars shown in Figure \ref{fig: cmds of contaminations}c, the vertical distributions (i) and (ii) can  be identified as Galactic halo and disk stars, respectively.   In our mock catalogue, nearby young disk stars ($< 5$ kpc) distribute vertically at $\oi>2$ and old halo stars populate at $\ogi \sim 0.3$, with both becoming redder at fainter magnitudes.  At a given magnitude, halo stars also become bluer with the increasing distance.  We note that most of stars bluer than this halo star sequence have high surface gravities ($\rm{\log (g) }>7$) in TRILEGAL indicating they are white dwarfs of the Milky Way. 

Figure \ref{fig: cmds of contaminations}d is the CMD of galaxies in the photometric catalogue of the Cosmic Assembly Near-infrared Deep Extragalactic Legacy Survey (CANDELS) in the Cosmic Evolution Survey (COSMOS) field with the colour-coded according to the photometric redshift.  We take the CANDELS/COSMOS catalogue from \citet{2017ApJS..228....7N} and show the Suprime-Cam g- and i-band magnitudes of sources selected to have SExtractor parameter $CLASS~STAR < 0.5$ (i.e. probable galaxies) in the HST/WFC F160W bandpass.  Although the CANDELS/COSMOS catalogue is deeper than our M81 catalogue, limited to $0.6~\sqdeg$ ($\sim13\%$ of M81 Field area) and the methods of photometry and star/galaxy separation are different from those employed in this study, the galaxy distribution in colour-magnitude space shows many similarities to the distribution (iii) in the M81 Field CMD.  Galaxies at $\oi>23$ and $\ogi<0.6$ have higher photometric redshifts ($z\sim2$), suggesting that the bluer objects in the Field CMD are unresolved high-z galaxies, consistent with the findings in \citet{2012MNRAS.419.1489B}. 

\begin{figure*}
\begin{center}
 \includegraphics[width=420pt]{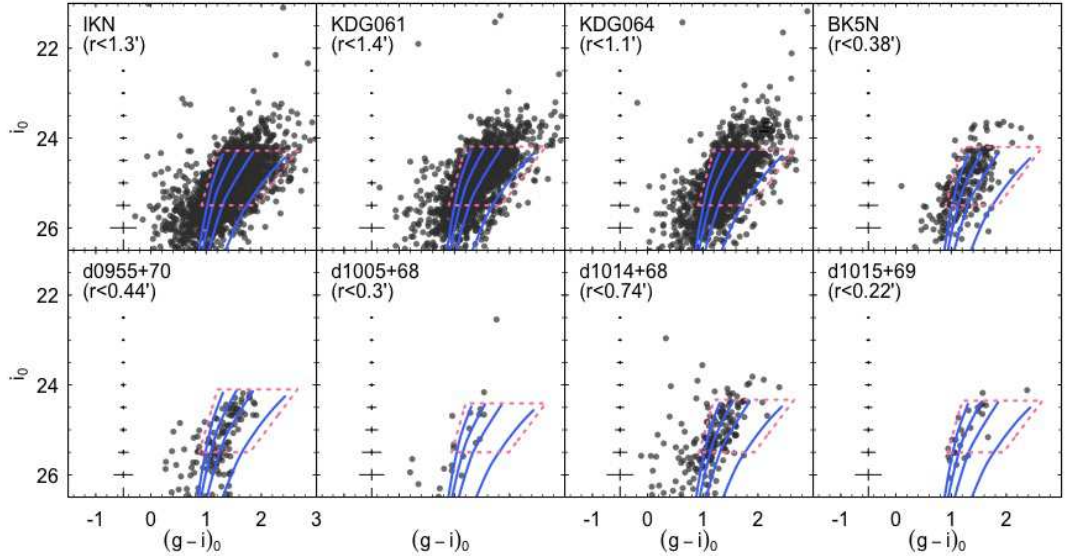}
 \vspace{5pt}
 \caption{De-reddened CMDs of stellar objects located within the half light radius of the respective dwarf galaxy.  The galaxy name and the limiting radius are indicated in each panel. The dotted polygon outlines the RGB selection box that was used for estimating structural parameters and metallicities. Theoretical PARSEC isochrones of a 10 Gyr old population with $\gmh=-2.2, -1.75, -1.3, -0.75$ are overlaid as magenta solid lines. The error bars show the photometric error at $\ogi=1$ in each galaxy as determined from the artificial star tests.}
 \label{fig: cmds of old dwarfs}
\end{center}
\end{figure*}

The number of foreground and background contaminants in our point source catalogue is significant, as can be seen in Figure \ref{fig: cmds of contaminations}, and these contaminants are likely to significantly outnumber genuine M81 Group stars in regions far from the main galaxies. Indeed, the M81 Field area we have defined should contain a large number of stars belonging to M81 and M82, since the stellar halos of M81 and M82 extend to beyond $4\times\rm{R}_{25}$ radii \citep{2015ApJ...809L...1O}.  In the following sections, we adopt colour and magnitude cuts that are designed optimise the selection of genuine M81 Group stars while limiting the number of foreground and background contaminants, and adjust them according to the distance of each galaxy.    

\section{Old Dwarf Galaxies in the M81 Group} \label{sec: old dwarfs}

Fields 1 to 4 include eight previously known dwarf spheroidal (dSph) galaxies of the M81 Group; IKN, KDG061, KDG064, BK5N, d0955+70, d1005+68, d1014+68, and d1015+69, as well as M81, M82 and NGC3077.  In this section, we derive, in a homogeneous manner, the structural parameters and the photometric metallicity distributions of these dSphs based on the HSC survey. 

\subsection{CMDs}
Figure \ref{fig: cmds of old dwarfs} shows the $\oi$ versus $\ogi$ CMDs of stellar objects lying within the half light radii ($\rh$) of the best-fit Sersic profile of galaxies, as derived in Section \ref{subsec: old dwarf structure}.  Theoretical PARSEC v1.2 isochrones of 
age 10 Gyr with $\gmh=-2.2, -1.75, -1.3, -0.75$ for the SDSS filter system \citep{2012MNRAS.427..127B} are overlaid,  adjusted to the distance of each galaxy using the distance moduli in the literature \citep{2002A&A...383..125K, 2006AJ....131.1361K, 2008MNRAS.384.1544S, 2013AJ....145..101K, 2013AJ....146..126C, 2017ApJ...843L...6S}, which is very close to the HSC filter system \citep{2018ApJ...853...29K, 2018PASJ...70...66K}.    
We have chosen to adopt literature distances (see Table \ref{tbl: old str2}), instead of deriving our own, since these are generally based on high precision photometry obtained with the  HST.  The error bars in each panel show the photometric error at $\ogi=1$ as derived from the artificial star tests.  The dotted polygons delineate the selection boxes of RGB stars for the estimation of structural parameters in Section \ref{subsec: old dwarf structure} and metallicities in Section \ref{subsec: old dwarf pop}.

Thanks to the photometric depth of our survey, we are able to resolve individual stars to at least 1.5 magnitude below the RGB tip (TRGB) of each dwarf galaxy.  In Figure \ref{fig: cmds of old dwarfs}, the dominant populations are seen to be metal-poor RGBs.  The
four brighter dwarfs, IKN, KDG061, KDG064 and BK5N, also show populations of asymptotic giant branch (AGB) stars brighter than 
the TRGB, indicating these systems contain intermediate age (0.5-5 Gyr) stars.  The small handful of objects blueward of the RGB selection boxes are likely to be high-z galaxies, as discussed in the previous section. 

\subsection{Structural Parameters} \label{subsec: old dwarf structure}

\begin{figure*}
\begin{center}
 \includegraphics[width=168pt]{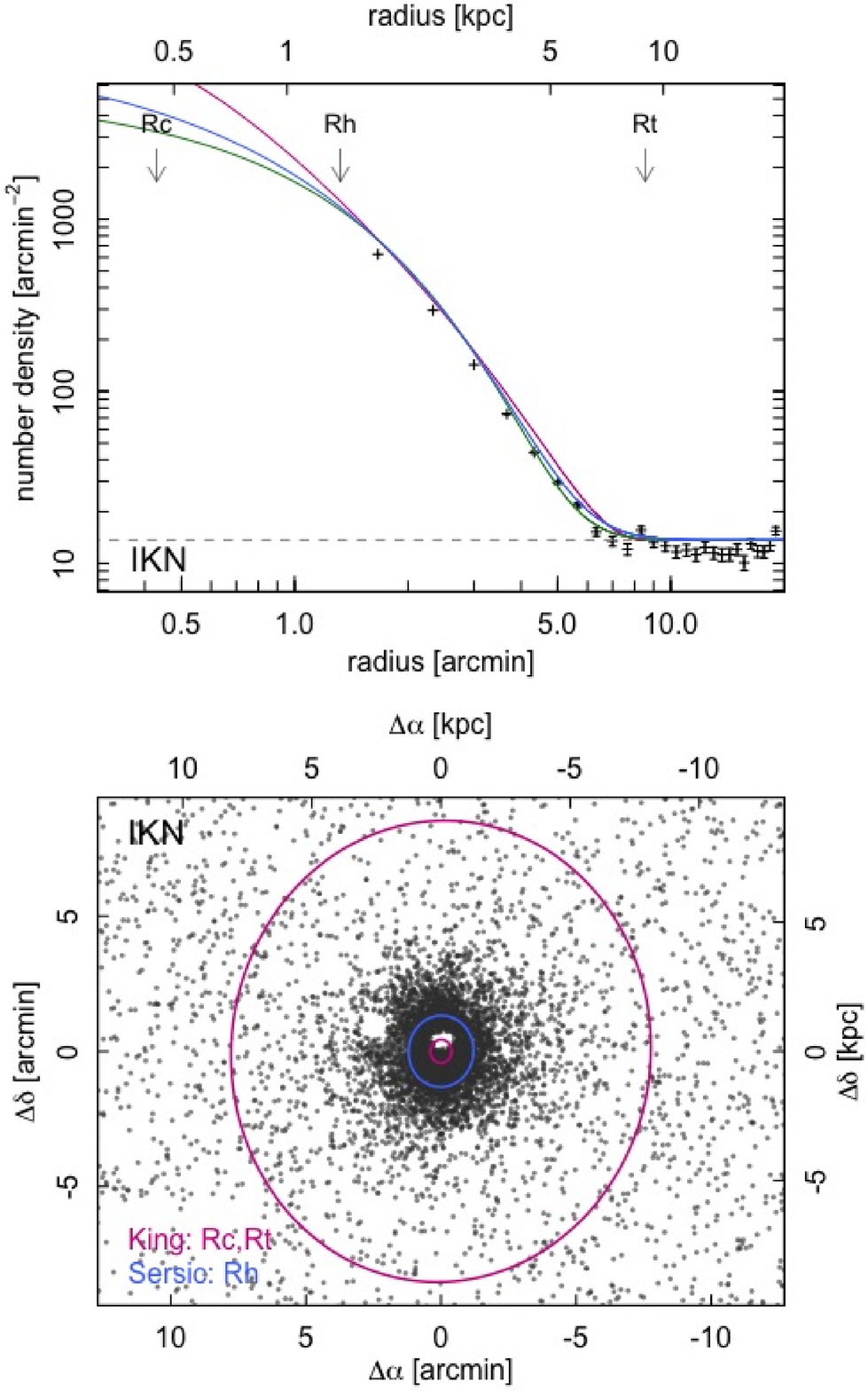}
 \includegraphics[width=168pt]{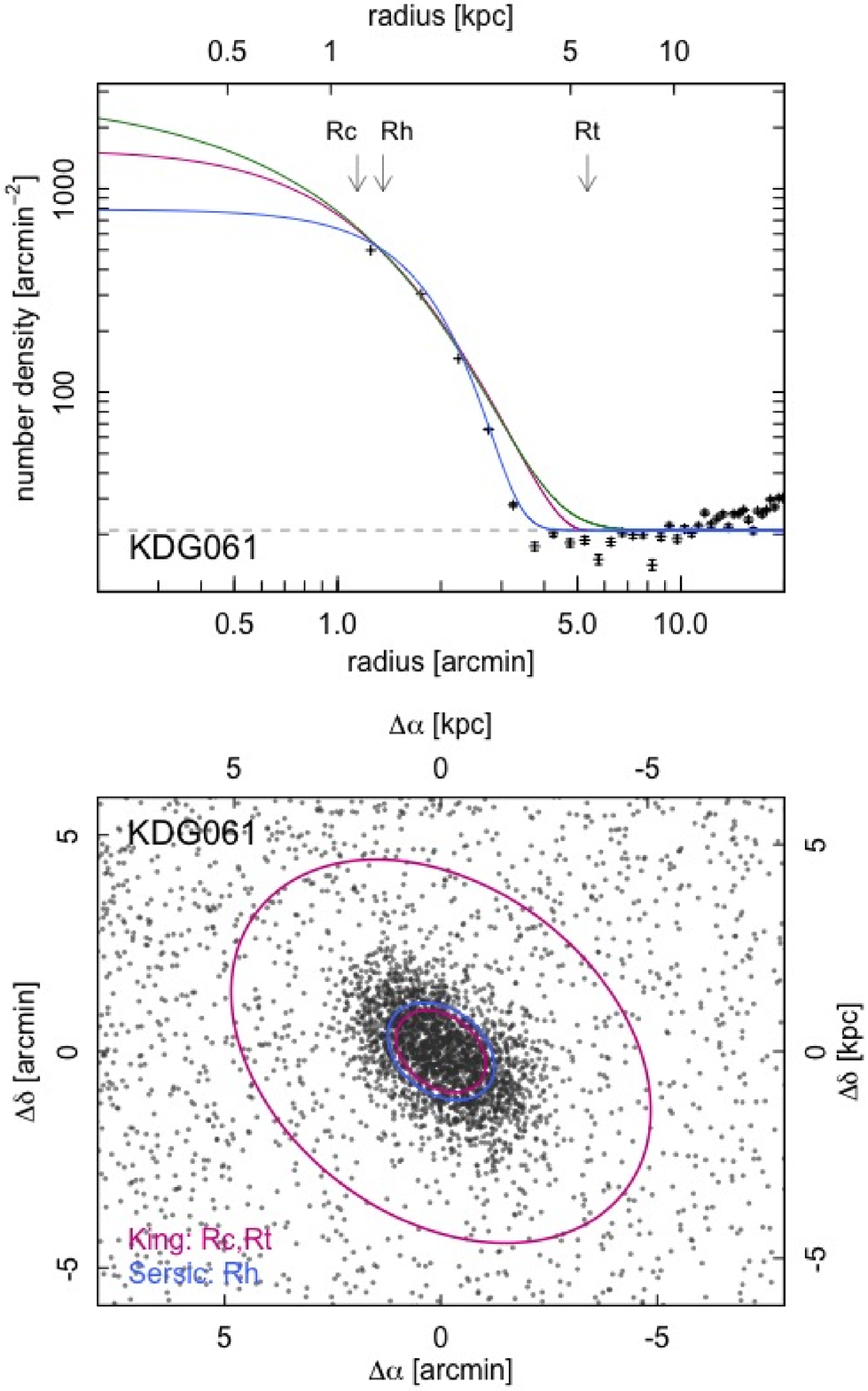}
 \includegraphics[width=168pt]{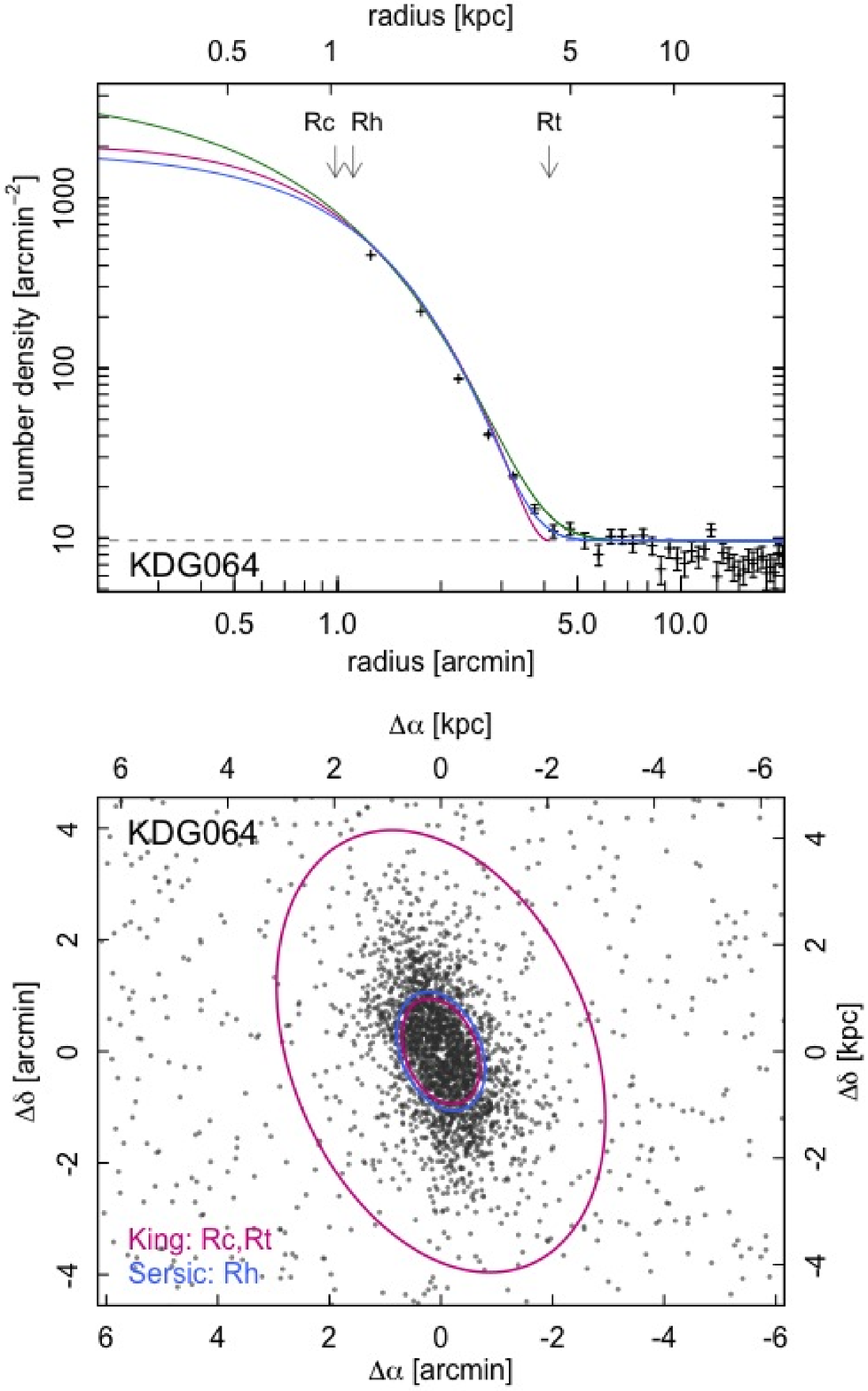}
 \vspace{5pt}
 \caption{Radial profiles and spatial distributions of RGB stars in IKN, KDG061, and KDG064.  {\it Upper panel:} Radial profile derived by calculating the average RGB number density (completeness-corrected) within elliptical annuli.  The best-fitting King, exponential, and Sersic profiles are overlaid as magenta, green, and blue lines, respectively.  The foreground/background contaminations are not subtracted and the contamination level is shown as a dashed line.  {\it Lower panel:} Spatial distribution of RGB stars.  The magenta lines and 
 the blue line represent $\rc$ and $\rt$ of the King profile, and $\rh$ of the Sersic profile, respectively. }
 \label{fig: map and profile 1}
\end{center}
\end{figure*}

\begin{figure*}
\begin{center}
 \includegraphics[width=168pt]{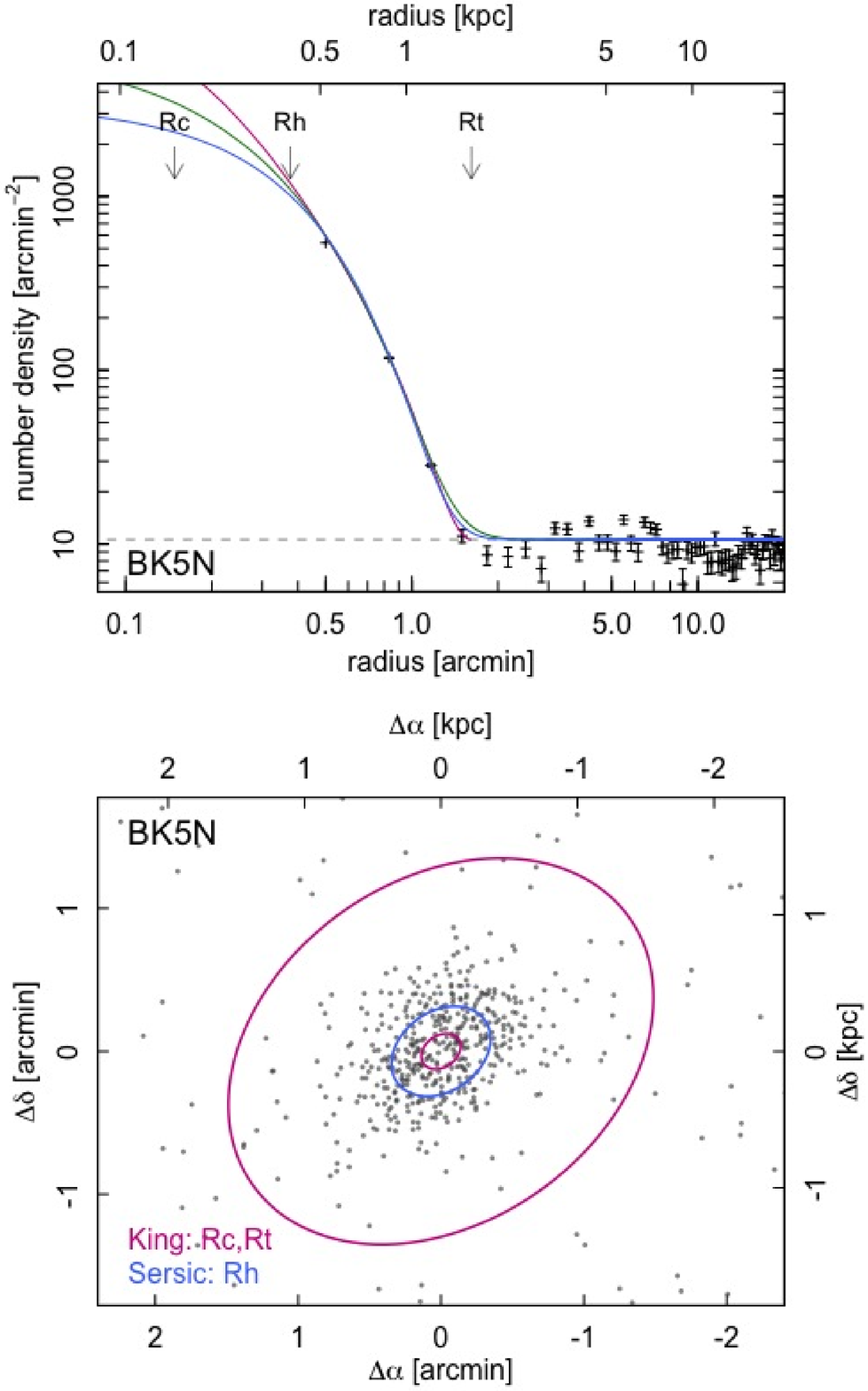}
 \includegraphics[width=168pt]{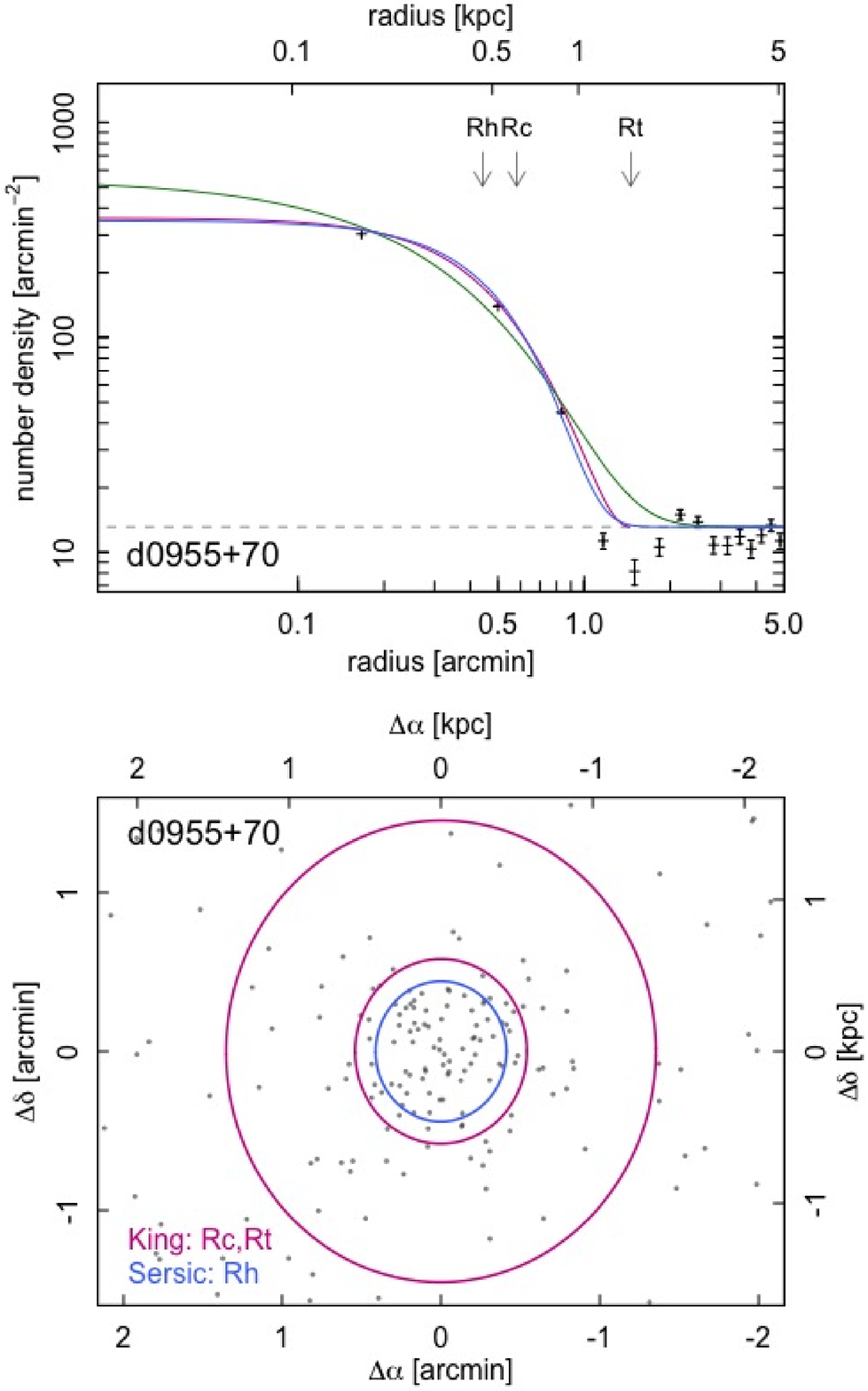}
 \includegraphics[width=168pt]{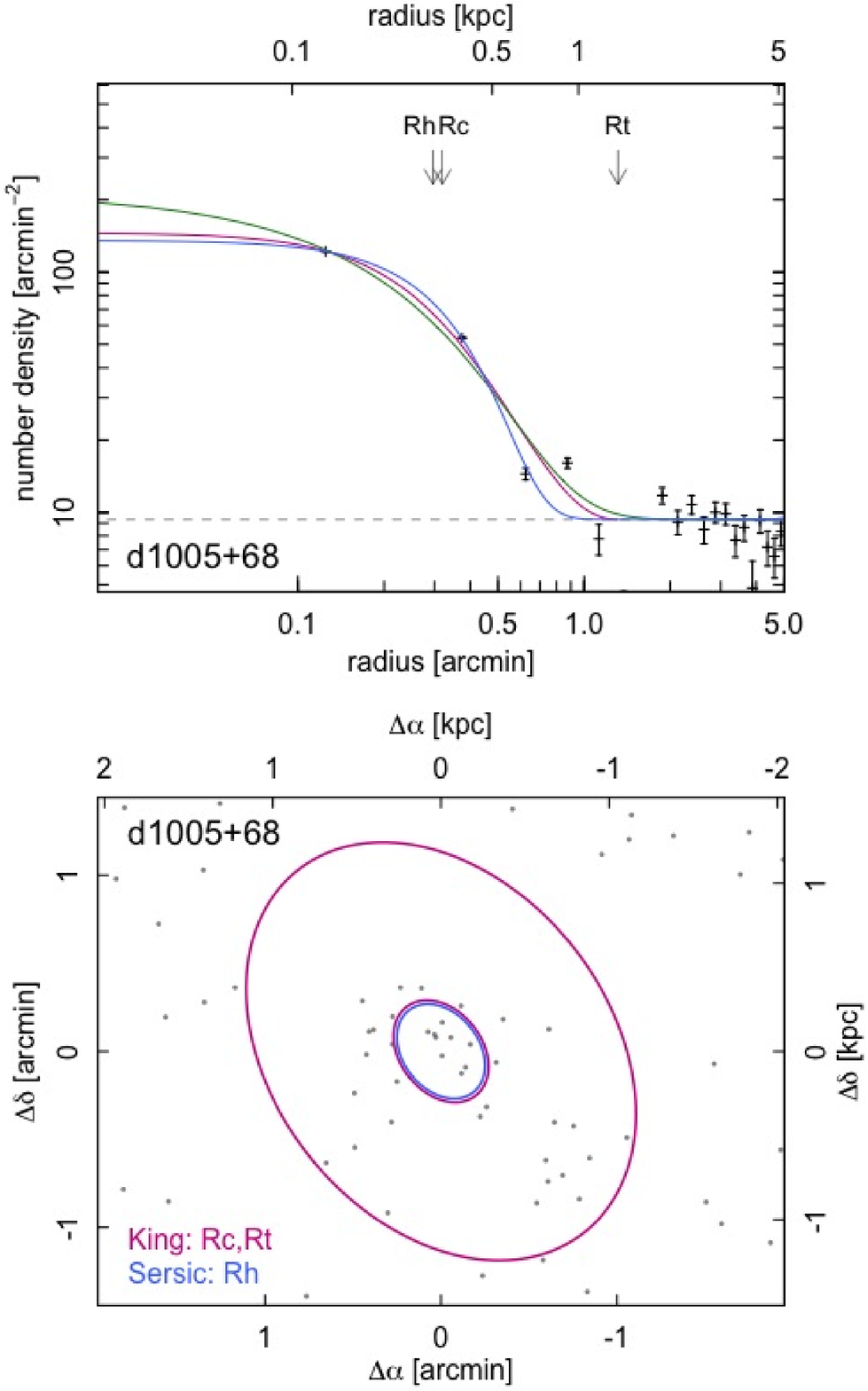}
 \vspace{5pt}
 \caption{Same as Figure \ref{fig: map and profile 1}, but for BK5N, d0955+70, and d1005+68. }
 \label{fig: map and profile 2}
\end{center}
\end{figure*}

\begin{figure*}
\begin{center}

 \includegraphics[width=168pt]{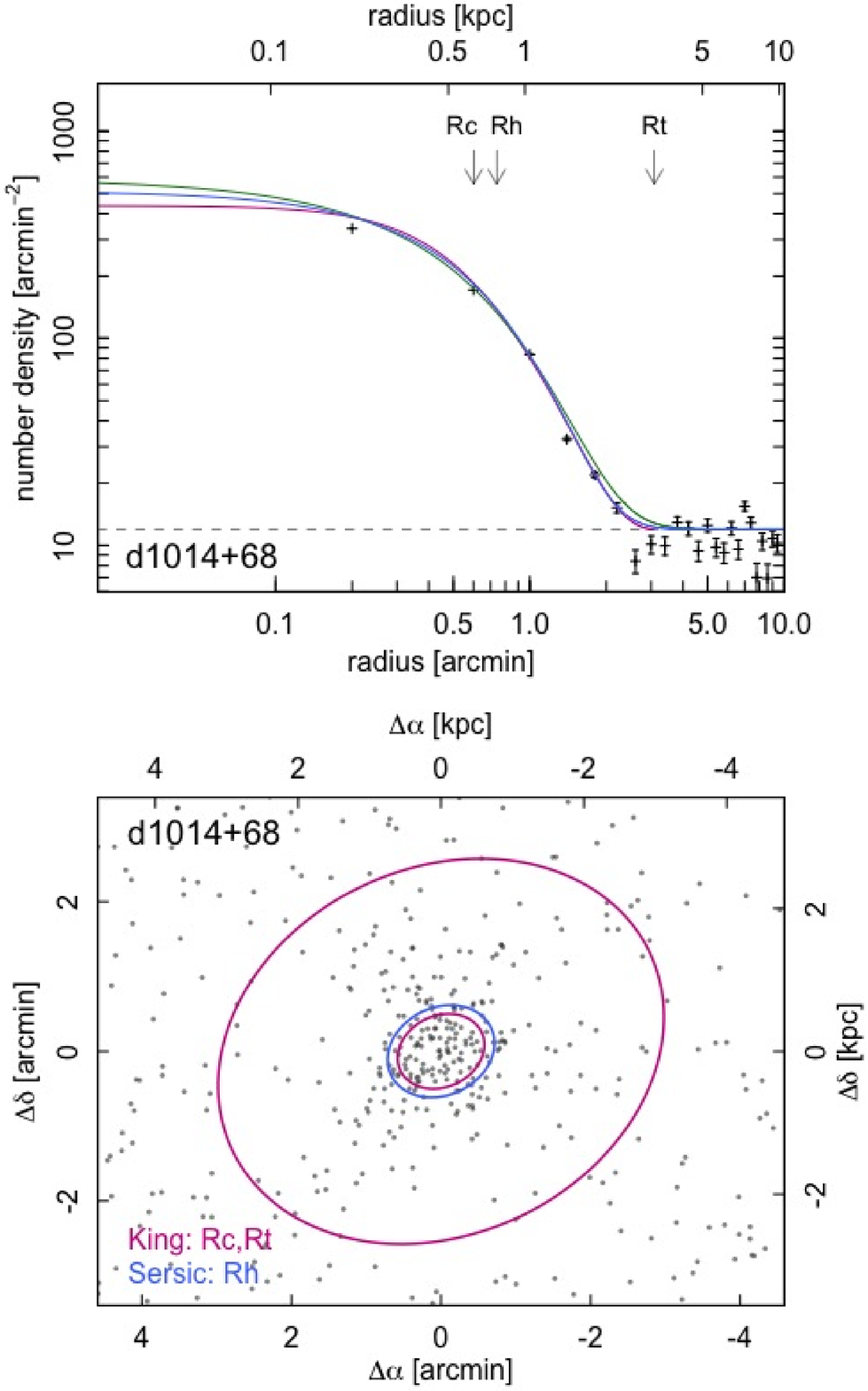}
 \includegraphics[width=168pt]{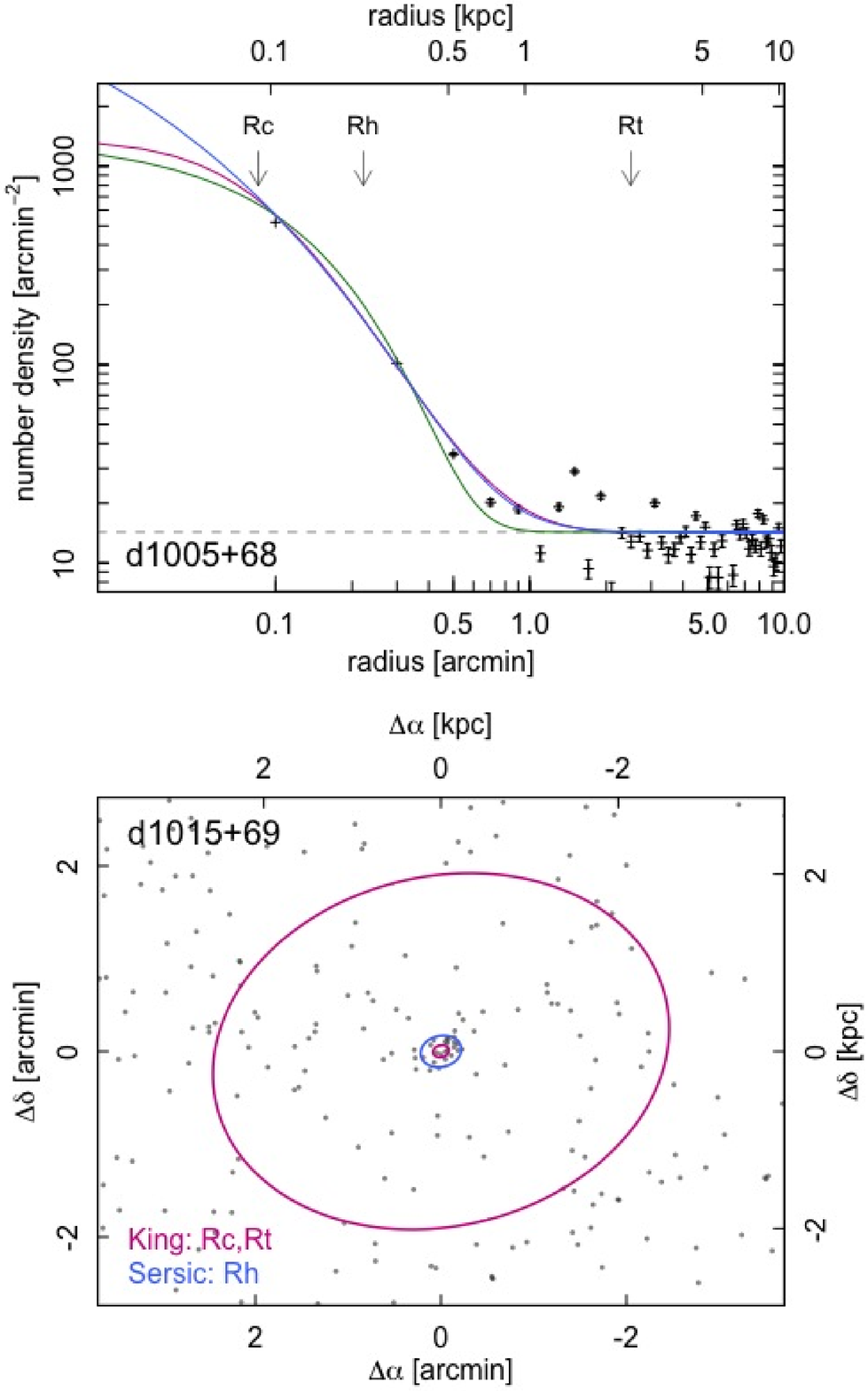}
 \vspace{5pt}
 \caption{Same as Figure \ref{fig: map and profile 1}, but for d1014+68 and d1015+69.}
 \label{fig: map and profile 3}
\end{center}
\end{figure*}

\begin{deluxetable*}{rccccccccc}
\tabletypesize{\scriptsize}
\tablecolumns{9}
\tablewidth{0pt}
\tablecaption{The structural parameters of the early-type dwarf galaxies \label{tbl: old str}}
\tablehead{
\colhead{Galaxy}  & \colhead{$\alpha$} & \colhead{$\delta$}   & \colhead{P.A.\tablenotemark{a}}  & \colhead{$\epsilon$\tablenotemark{b}}   &
\colhead{$\rc$\tablenotemark{c}} & \colhead{$\rt$\tablenotemark{d}} & \colhead{$\re$\tablenotemark{e}} &   
\colhead{n\tablenotemark{f}}   & \colhead{$\rh$\tablenotemark{g}}\\
 & (J2000) & (J2000) & deg & & arcmin & arcmin & arcmin & & arcmin} 
\startdata
    IKN & $10^{h}08^{m}05^{s}.0$ & $+68\arcdeg25\arcmin16\arcsec.1$ & $174.6$ & $0.10$ & $0.4\pm0.1$   & $8.5\pm0.4$ & $0.85\pm0.01$ & $1.24\pm0.03$ & $1.32\pm0.01$\\
 KDG061 & $09^{h}57^{m}03^{s}.2$ & $+68\arcdeg35\arcmin34\arcsec.4$ &  $53.1$ & $0.28$ & $1.1\pm0.2$   & $5.3\pm0.4$ & $0.74\pm0.04$& $0.40\pm0.01$ & $1.36\pm0.01$\\
 KDG064 & $10^{h}07^{m}02^{s}.0$ & $+67\arcdeg49\arcmin43\arcsec.4$ &  $22.3$ & $0.35$ & $1.0\pm0.2$   & $4.1\pm0.3$ & $0.59\pm0.01$ & $0.70\pm0.07$ & $1.11\pm0.03$\\
   BK5N & $10^{h}04^{m}41^{s}.6$ & $+68\arcdeg15\arcmin24\arcsec.8$ & $125.4$ & $0.26$ & $0.15\pm0.04$ & $1.62\pm0.1$ & $0.20\pm0.01$ &$0.77\pm0.08$ & $0.38\pm0.02$\\
d0955+70& $09^{h}55^{m}14^{s}.3$ & $+70\arcdeg24\arcmin24\arcsec.2$ & $178.9$ & $0.07$ & $0.58\pm0.07$ & $1.5\pm0.1$ & $0.31\pm0.04$ & $0.51\pm0.07$ & $0.44\pm0.02$\\
d1005+68& $10^{h}05^{m}32^{s}.3$ & $+68\arcdeg14\arcmin17\arcsec.3$ &  $38.4$ & $0.27$ & $0.3\pm0.2$   & $1.3\pm0.9$ & $0.22\pm0.04$ & $0.5\pm0.2$   & $0.30\pm0.05$\\
d1014+68& $10^{h}14^{m}54^{s}.7$ & $+68\arcdeg45\arcmin35\arcsec.9$ & $115.2$ & $0.20$ & $0.60\pm0.05$ & $3.1\pm0.5$ & $0.48\pm0.02$ & $0.9\pm0.1$   & $0.74\pm0.04$\\
d1015+69& $10^{h}15^{m}06^{s}.8$ & $+69\arcdeg02\arcmin11\arcsec.6$ & $103.5$ & $0.24$ & $0.09\pm0.03$ & $2.5\pm2.0$ & $0.11\pm0.01$ & $2.5\pm1.5$   & $0.22\pm0.04$\\
\enddata
\tablecomments{(a) Position angle from north to east. (b) Ellipticity $\epsilon=1-\rm{b/a}$ where $\rm{b/a}$ is the axis ratio of a galaxy. (c) Core radius of the King profile. (d) Tidal radius of the King profile. (e) Scalelength of the exponential profile. (f) Sersic index. (g) Half light radius of the Sersic profile. }
\end{deluxetable*}

\begin{deluxetable*}{rcccccccccc}
\tabletypesize{\scriptsize}
\tablecaption{The properties of the early-type dwarf galaxies \label{tbl: old str2}}
\tablehead{
\colhead{Galaxy} &\colhead{type} &\colhead{Distance} & \colhead{$\rc$} & \colhead{$\rt$} &\colhead{$c$\tablenotemark{a}} &\colhead{$\rh$} &\colhead{$\mv$} &\colhead{$\vi$} &\colhead{$\langle\gmh_{w}\rangle$\tablenotemark{b}} &\colhead{$\sigma_{\gmh}$\tablenotemark{c}}\\
 & & Mpc  & kpc & kpc & & kpc & mag & mag & dex & dex }
\startdata
IKN            & dE    & $3.75\pm0.45 \tablenotemark{(1)}$ & $0.5\pm0.2$  & $9.3\pm0.4$ & 1.27 & $1.44\pm0.02$ & $-14.29\pm0.49$ & $1.37\pm0.22$ & $-1.17\pm0.23$ & 0.39 \\
KDG061    & dSph & $3.60\pm0.43 \tablenotemark{(1)}$ & $1.2\pm0.2$  & $5.6\pm0.4$ & 0.67 & $1.42\pm0.01$ & $-13.38\pm0.52$ & $1.36\pm0.22$ & $-1.29\pm0.21$ & 0.38 \\
KDG064    & dSph & $3.70\pm0.44 \tablenotemark{(1)}$ & $1.1\pm0.2$  & $4.4\pm0.3$ & 0.60 & $1.19\pm0.04$ & $-13.31\pm0.55$ & $1.34\pm0.23$ & $-1.32\pm0.30$ & 0.37 \\
BK5N        & dSph & $3.78\pm0.45 \tablenotemark{(1)}$ & $0.16\pm0.04$ & $1.8\pm0.03$ & 1.05 & $0.41\pm0.02$ & $-11.23\pm0.58$ & $1.26\pm0.23$ &  $-1.60\pm0.34$ & 0.32 \\
d0955+70 & dSph & $3.45~^{+~0.60}_{-~0.46} \tablenotemark{(2)}$ & $0.6\pm0.1$ & $1.5\pm0.1$ & 0.40 & $0.44\pm0.02$ & $-10.36\pm0.47$ & $1.32\pm0.35$ & $-1.55\pm0.23$ & 0.27 \\
d1005+68 & dSph & $3.98~^{+~0.38}_{-~0.43} \tablenotemark{(3)}$ & $0.4\pm0.2$   & $1.5\pm1.1$ & 0.57 & $0.34\pm0.05$ & $-8.14\pm0.55$ & $1.25\pm0.26$ & $-1.79\pm0.37$ & 0.28 \\
d1014+68 & dSph & $3.84\pm0.33 \tablenotemark{(2)} $ & $0.67\pm0.06$ & $3.4\pm0.6$ & 0.70 & $0.82\pm0.05$ & $-11.28\pm0.48$ & $1.28\pm0.24$ &  $-1.50\pm0.27$ & 0.33 \\
d1015+69 & dSph & $3.87~^{+~0.26}_{-~0.21} \tablenotemark{(2)}$ & $0.10\pm0.03$ & $2.8\pm2.2$ & 1.45 & $0.25\pm0.05$ & $-9.2\pm0.41$ & $1.26\pm0.20$ & $-1.61\pm0.28$ & 0.31 \\
\enddata
\tablecomments{(a) Concentration parameter $c=\log(\rt/\rc)$ (b) The weighted mean metallicity (c) The metallicity dispersion. }
\tablerefs{ (1) \citet{2013AJ....145..101K}, \ (2) \citet{2013AJ....146..126C}, (3) \citet{2017ApJ...843L...6S} }
\end{deluxetable*}

We compute the structural parameters of the eight dwarf galaxies using the density-weighted first and second moments of the RGB star spatial distribution. The selection criteria for the RGB stars is shown as the dotted polygon in each panel of Figure \ref{fig: cmds of old dwarfs}.  We define the RGB box to include stars in the colour interval sandwiched by the $\gmh=-2.2$ and $\gmh=-0.7$ isochrones, convolved with the photometric errors.  This interval is deemed to be sufficiently wide to capture the expected populations, while avoiding areas that are densely populated by contaminants. 

We find values for the centroids, average ellipticities and position angles that are consistent with those from previous studies except for the declination of IKN, which we find to be $1.3\arcmin$ north of the value listed in the NASA/IPAC Extragalactic Database (NED)\footnote{http://ned.ipac.caltech.edu/}. Since IKN overlaps with two bright foreground stars located at about $0.5\arcmin$ and $2.7\arcmin$ north of the IKN center, the scattered light from these stars may have prevented accurate estimation of
its center (see two holes in the lower left panel of Figure \ref{fig: map and profile 1}).  To reduce these effects, we populated the areas
affected by these bright stars with stars drawn from a similarly-sized axially-symmetric position before estimating the structural parameters of IKN.  The resulting coordinate is consistent with that in \citet{2015A&A...581A..84T} which derived the center from an HST/ACS stellar catalogue covering the southern half of IKN. 
 
Table \ref{tbl: old str} lists the structural parameters of the eight dwarf galaxies, which are derived from
fitting the standard King, exponential, and Sersic profiles via least-squares minimisation to the completeness-corrected RGB star counts.  The completeness of a given RGB star is inferred from comparing its magnitude and colour to a modified sigmoid function fit to the point source detection ratio as functions of the magnitude and the color, as given by the artificial star tests.  
The radial star count profiles are then constructed by calculating the average number density of RGB stars in a series of elliptical annuli, defined using the structural parameters derived above. 
The field contamination level at the location of each galaxy is calculated firstly as an average of number density in a certain radius range of outer region by eye. Then we fit the profiles and calculate the contamination level again as a trimmed mean of number density between radii of $\rt < r < 2*\rt$ for relatively large IKN, KDG061, KDG064, and $\rt < r < 4*\rt$ for small galaxies.  Finally the profiles are fitted using these values.  Figures \ref{fig: map and profile 1}, \ref{fig: map and profile 2} and \ref{fig: map and profile 3} show the radial profiles and spatial distributions of the RGB stars in the eight galaxies.  Remarkably, this is
the first estimation of structural parameters for IKN, even although it is one of the most well-studied satellites in the M81 Group.  Again,
this is probably due to the presence of the bright foreground stars which hinder any simple analysis. The exponential scalelengths $\re$ of KDG061 and KDG064, and the Sersic radii $\rh$ of d0955+70, d1014+68, d1015+69 are all larger than the previous estimations ($\re=0.48\arcmin, 0.28\arcmin$, and $\rh=0.31\arcmin, 0.33\arcmin, 0.14\arcmin$, respectively) based on integrated light surface photometry \citep{2008MNRAS.384.1544S, 2013AJ....146..126C}.  The Sersic radius $\rh$ of BK5N is consistent with that ($\rh=0.4\arcmin$) of \citet{1998AJ....115..535C} while the Sersic index is much smaller than their value ($n=1.81$), also derived from
integrated light photometry.   While these differences are likely due to the greater stellar extents of these systems as mapped in our survey, they could also reflect issues with our completeness correction in the central regions of the brightest systems.  In
the case of d1005+68, the $\rh$ we find is much larger than that ($\rh=0.16\arcmin$) estimated by \citet{2017ApJ...843L...6S}, also
using resolved stars.  This is likely due to their assumption of a circular distribution; if we also assume the ellipticity of d1005+68 is $\epsilon=0$, we obtain a similar value of $\rh=0.19\arcmin$.

The radial profiles shown in Figures \ref{fig: map and profile 1} to \ref{fig: map and profile 3} are well described by the Sersic and King profiles.  It is curious that the number density of RGB stars in  KDG061 gradually increases beyond the tidal radius.  Since KDG061 is the nearest dSph to M81 having a projected distance $\rm{D}_{\rm{M81,pro}}=31$ kpc, this is likely due to contamination by M81 (and perhaps also NGC3077) halo stars.  BK5N also has a slight excess of stars at around 5 kpc, which is likely due to stars belonging to another dSph, d1005+68, located at 5 kpc from BK5N center.  The faintest two galaxies, d1005+68 and d1015+69, show small over-densities at a radial distance of roughly 2 kpc, however there is no visible structure in the RGB maps and these enhancements may just reflect noise.  Overall, no obvious tidal features, such as S-shaped structures, are found in these galaxies, although some deviations from the King profile can be seen in Figures \ref{fig: map and profile 1} to \ref{fig: map and profile 3}.  We return to this topic in Section \ref{subsec: tidal effect}.

Using stars that lie within the best-fit Sersic radius $\rh$, we estimate the total absolute magnitudes $\rm{M}_g$ and $\rm{M}_i$ of
the dwarf galaxies.  The total flux of RGB stars within $\rh$ is doubled and then corrected to account for the photometric completeness, contaminants, and stars fainter than the RGB selection box.  The V-band absolute magnitudes $\mv$, $\vi$ colours and their errors are calculated from
transforming $\rm{M}_g$ and $\rm{M}_i$ \citep{2006A&A...460..339J} and are listed in Table \ref{tbl: old str2}.  The errors include the photometric error, the distance error, and the uncertainties of the adopted corrections.  While the $\mv$ of KDG061, KDG064, BK5N and d1005+68 are all in good agreement with the previous estimations \citep{1998AJ....115..535C, 2000A&A...363..117K, 2017ApJ...843L...6S}, we find $\mv=-14.29$ for IKN which is almost 3 magnitudes brighter than the previous estimation of  \citet{2009MNRAS.392..879G} ($\mv=-11.51$) derived from the total B-band magnitude and assuming an average $(B-V)=0.45$ and $(V-I)=0.7$.  To estimate the luminosity of IKN, we again populated the areas
affected by the bright stars with stars drawn from similarly-sized axially-symmetric positions.  Since our estimation is less affected by  the PSF halos of these bright foreground stars than integrated light studies, we conclude our measurement is more reliable than the previous estimation.  This elevates IKN to being the brightest dwarf satellite in the M81 Group and lying in the luminosity-size realm of systems typically classified as dwarf ellipticals.

\subsection{Stellar Populations and Metallicity} \label{subsec: old dwarf pop} 
\begin{figure*}
\begin{center}
 \includegraphics[width=400pt]{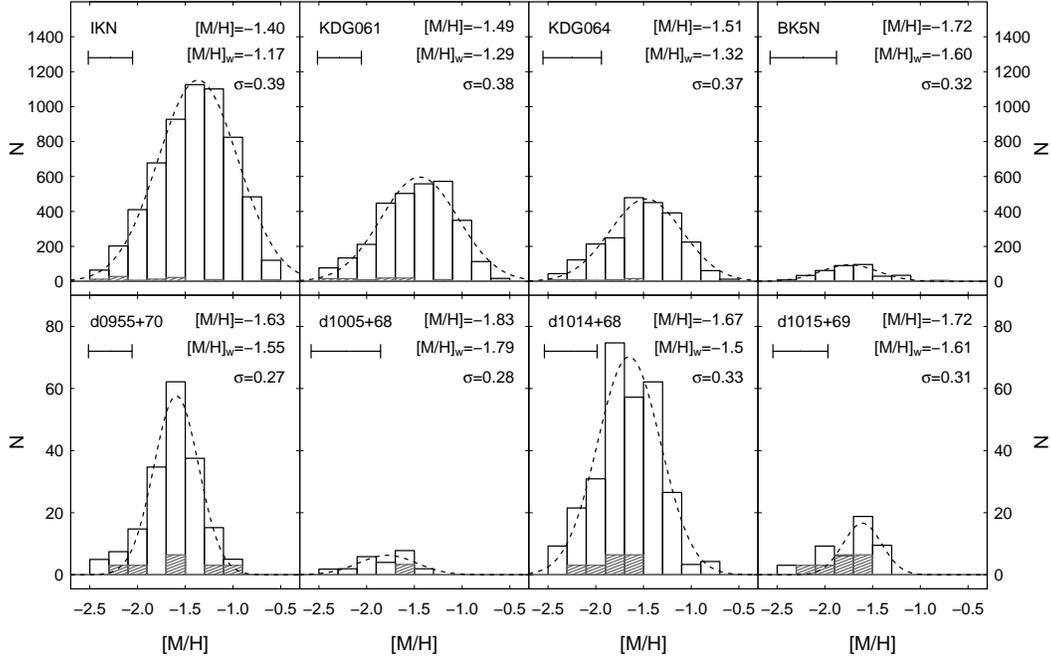}
 \vspace{5pt}
 \caption{Metallicity distribution of early-type dwarf galaxies. The dashed lines show single gaussian fits to the distributions and
 the grey shaded histograms are the distributions of the field contaminants. The galaxy name, mean metallicity, weighted mean metallicity and dispersion are indicated in each panel.}
 \label{fig: MDF of 8 old dwarfs}
\end{center}
\end{figure*}

Figure \ref{fig: MDF of 8 old dwarfs} shows the metallicity distribution functions (MDFs) of the dwarf galaxies as derived from the completeness-corrected colour distributions of bright RGB stars ($\oi<25$).  These stars are selected according to the criteria in Figure \ref{fig: cmds of old dwarfs} and lying within $2\times\rh$ of the galaxy center.  The photometric metallicity of each RGB star is derived from the linear interpolation of isochrones of $\gmh=-2.2$ to $\gmh=-0.7$ with a fixed age of 10 Gyr.  Note that we choose a 10 Gyr population so as to cover stellar populations in dwarf galaxies having wide luminosity range from $\mv=-14$ to $-8$. The uncertainty
introduced by  assuming a constant old age on the MDFs is quite small as shown in \citet{2010A&A...521A..43L}. The difference in the  resulting metallicity among the cases of fixed ages of 12.5, 10.5, and 8.5 Gyr in their study is less than 0.2 dex, and the shape of MDFs does not change significantly.   

We use RGB stars brighter than $\oi=25$ to reduce the effects of spatial variance on the photometric error and completeness, and to keep wide spacing between isochrones in the $g-i$ colour.  The error bar in each panel indicates the mean error estimated by performing a Monte Carlo simulation with N$=2000$ for each RGB star within its photometric uncertainty. Each star is randomly resampled from a Gaussian distribution with width equal to the photometric error and the metallicity is re-determined.  The standard deviation of the resultant distribution is adopted as the metallicity uncertainty for each RGB star.  Due to the narrow range of RGB colour between metal-poor isochrones, the estimated uncertainties increase with the decreasing metallicity.  The mean metallicity corrected for this effect is shown as the error-weighted mean value in each panel of Figure \ref{fig: MDF of 8 old dwarfs} and listed in Table \ref{tbl: old str2}.  The dashed line represents the a single gaussian fit to the MDF.  Foreground and background contamination is estimated using point sources drawn from the area outside of $\rt$ of each galaxy, and excluding regions populated by other galaxies or overdensities.  As can be seen in the grey shaded histograms of Figure \ref{fig: MDF of 8 old dwarfs}, the effect of contamination on the resultant MDFs is generally negligible. 
 
The estimated metallicity of IKN is lower than the previous estimation ($\langle\feh\rangle_w=-1.08$) using the HST/ACS data \citep{2010A&A...521A..43L}.  Although their HST photometry reached a deeper magnitude than ours, they were limited to a small FOV and hence likely missed more metal-poor stars in the outer regions of the galaxy. On the other hand,  our HSC photometry will definitely miss some (presumably more metal-rich) stars within the central half arcminute area due to the crowding.  The weighted-mean metallicities of KDG061 and KDG064 are higher than those reported in \citet{2010A&A...521A..43L}  ($\langle\feh\rangle_w=-1.49$ and $-1.57$, respectively), but the differences are within the uncertainties.  The non-weighted mean metallicities of BK5N and d1005+68 are consistent with those values $\feh=-1.7\pm0.4$ of \citet{2000A&A...363..117K} and $\feh=-1.9\pm0.24$ of \citet{2017ApJ...843L...6S}, respectively. 

The MDFs in Figure \ref{fig: MDF of 8 old dwarfs} are generally well described by a single gaussian.  There is a slight excess of metal-poor stars in d0955+70, d1014+68 and d1015+69, but this might simply reflect the fact that our metallicity estimation is limited to $\gmh>-2.5$ due to the available isochrones.  The overall symmetric appearance of the MDFs implies that there was no sudden truncation of star formation in the past, which would be expected to cause sharp drops on the metal-rich side \citep{1987A&A...188...13Y}.  Although the resulting MDFs do not contain particularly apparent metal-poor tails, the overall shapes resemble the combined MDF of similarily luminous Galactic dSphs shown in \citet{2013ApJ...779..102K}, which are well reproduced by an infall model.  

\section{The dwarf galaxy candidates}
\label{new dwarf}
\begin{figure}
\begin{center}
 \includegraphics[width=250pt]{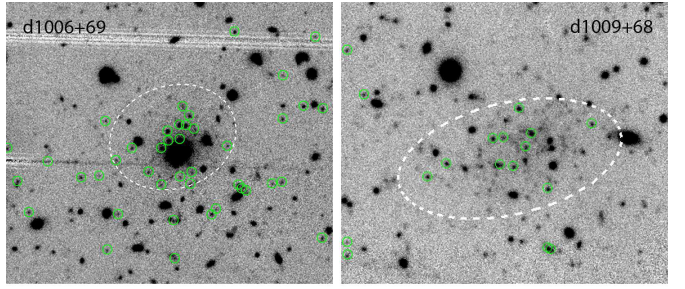}
 \caption{HSC i-band images covering about $1.1\arcmin \times 0.95\arcmin$ of d1006+69 (left) and d1009+68 (right).  The dashed white ellipses correspond to the half-light radii and green circles are member RGB stars brighter than $\oi=26.5$.  }
 \label{fig: uko img}
\end{center}
\end{figure}

\begin{figure}
\begin{center}
 \includegraphics[width=265pt]{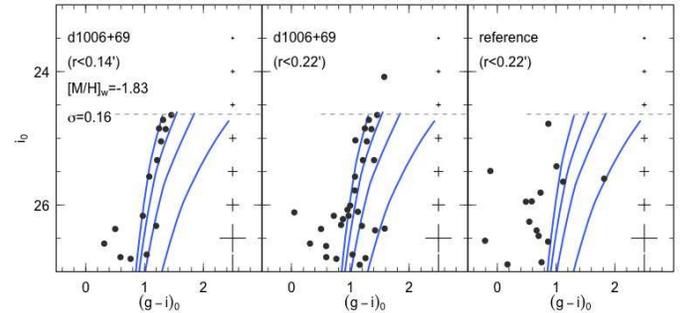}
 \caption{The CMDs of the dwarf galaxy candidate d1006+69 and a nearby reference reference field.  The panels show stellar objects lying within the core radius ({\it Left}), within the Sersic $\rh$ radius ({\it Middle}), and in a field the same size as that shown in the middle panel but located 
 $5\arcmin$ away from the galaxy center ({\it Right}).  The horizontal dashed line represents the TRGB and the overlaid isochrones are the same as those shown in Figure \ref{fig: cmds of old dwarfs}. The error bars show the photometric error at $\ogi=1$.}
 \label{fig: uko01 cmd}
\end{center}
\end{figure}

\begin{figure}
\begin{center}
 \includegraphics[width=265pt]{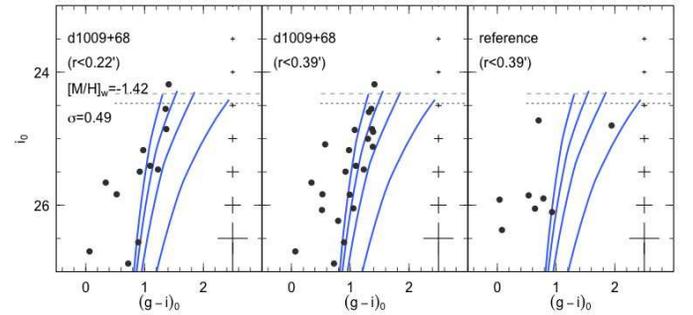}
 \caption{Same as Figure \ref{fig: uko01 cmd}, but for d1009+68.  The horizontal dotted line is the transition detected by a Sobel filter. }
 \label{fig: uko02 cmd}
\end{center}
\end{figure}

We conducted a visual search for new stellar overdensities in the M81 Group RGB map presented here.  
Most previous searches for dwarf galaxies in the M81 Group have not had the sensitivity to detect very faint and low surface brightness systems \citep{2009AJ....137.3009C, 2013AJ....146..126C}.  An exception to this is the recent study by \citet{2017ApJ...843L...6S} who also used HSC to image an area around M81 which led to the discovery of a new satellite.  Using four HSC pointings around M81, we examined both the raw star count map and a gaussian-convolved version, and detected several candidate overdensities.  Most of these were dismissed as candidate dwarf galaxies through visual inspection of the pixel data and/or from examining CMDs, because they do not have a significantly resolved RGB.  However, two prominent overdensities, which we name d1006+69 and d1009+68, show convincing RGB sequences on the CMDs (see Figure \ref{fig: uko01 cmd} and \ref{fig: uko02 cmd}).  Figure \ref{fig: uko img} shows portions of the HSC $i$-band image showing these objects.  It can be seen
that d1006+69 projects close to the galaxy SDSSJ100655.25+695413.4, which has a photometric redshift of z$\sim0.319$.  The stars we identify as members of d1006+69 are too bright to be misidentified globular clusters associated with this background system and furthermore they do not distribute symmetrically around this galaxy. 

\begin{figure}
\begin{center}
 \includegraphics[width=240pt]{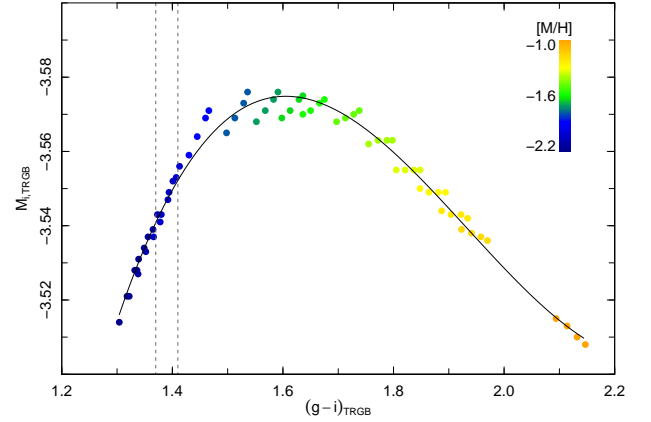}
 \caption{The $i$-band absolute magnitude of the TRGB in old (10-13 Gyr), metal-poor populations as a function of $(g-i)$ colour of TRGB, derived from the PARSEC v1.2 isochrones.  The colour of each point represents [M/H] and the black solid line shows the best fit polynomial equation \ref{eq:1}. The vertical dashed lines indicate the measured $\ogi$ colour of d1006+69 (left) and d1009+68 (right), respectively.}
 \label{fig: TRGB}
\end{center}
\end{figure}

We derive the distances to d1006+69 and d1009+68 using the TRGB method \citep{1993ApJ...417..553L}.  Stars of $0.5 < \ogi < 2.0$ which lie within $25\arcsec$ of the centers of the overdensities are used to derive the $i$-band luminosity function (LF).  We correct for foreground and background contamination using the LF of point sources in regions located $5\arcmin$ away from each candidate.  We apply a Sobel filter to detect the sharp transition at $\oi=24.64\pm0.08$ for d1006+69, and a relatively weak signal at $\oi=24.47\pm0.12$ for d1009+68, respectively.   At the center of d1009+68, one star ($\oi=24.18\pm0.02$, $\ogi=1.41\pm0.04$) exists brighter than this magnitude (see dotted lines of Figure \ref{fig: uko02 cmd}).  So we adopt the mean of these two values $\oi=24.325$ as the TRGB magnitude of d1009+68 and the difference 0.29 mag as the error.  Then we assume the colour of TRGB as $\ogi=1.37 \pm 0.12$ for d1006+69 and $\ogi=1.36 \pm 0.07$ for d1009+68, which are the average colours of the three stars nearest to the detected edge of the LF ($\oi=24.65, 24.72, 24.85$ for d1006+69 and $\oi=24.18, 24.55, 24.60$ for d1009+68).  Figure \ref{fig: TRGB} shows how the $i$-band absolute magnitude of the TRGB varies with $g-i$ colour according to PARSEC v1.2 old metal poor ($10-13$ Gyr, $\gmh < -1.0$) isochrones on the SDSS filter system.  We fit a polynomial 
relationship to these datapoints:
   
\begin{eqnarray}
\rm{M}_{i,\rm{TRGB}} = &-& 0.504\times\gitrgb^3 + 2.924\times\gitrgb^2 \nonumber \\
&-& 5.490\times\gitrgb - 0.211 \pm 0.008
\label{eq:1}
\end{eqnarray}

Given the measured $\ogi$ colours, we use this equation to calculate the expected $\rm{M}_{i,\rm{TRGB}}$ to be $-3.54\pm0.06$ for d1006+69 and $-3.54\pm0.03$ for d1009+68, respectively.   Finally, combining these values with the observed 
$i_{0,\rm{TRGB}}$, we calculate distance moduli as $\mm=28.18 \pm 0.10$ for d1006+69 which corresponds to $4.33\pm0.20$ Mpc, and $\mm=27.86 \pm 0.29$ for d1009+68 which corresponds to $3.7\pm0.5$ Mpc.  The errors include both the photometric errors as derived from artificial star tests and the uncertainties in the TRGB calibration.   

\begin{figure}
\begin{center}
 \includegraphics[width=170pt]{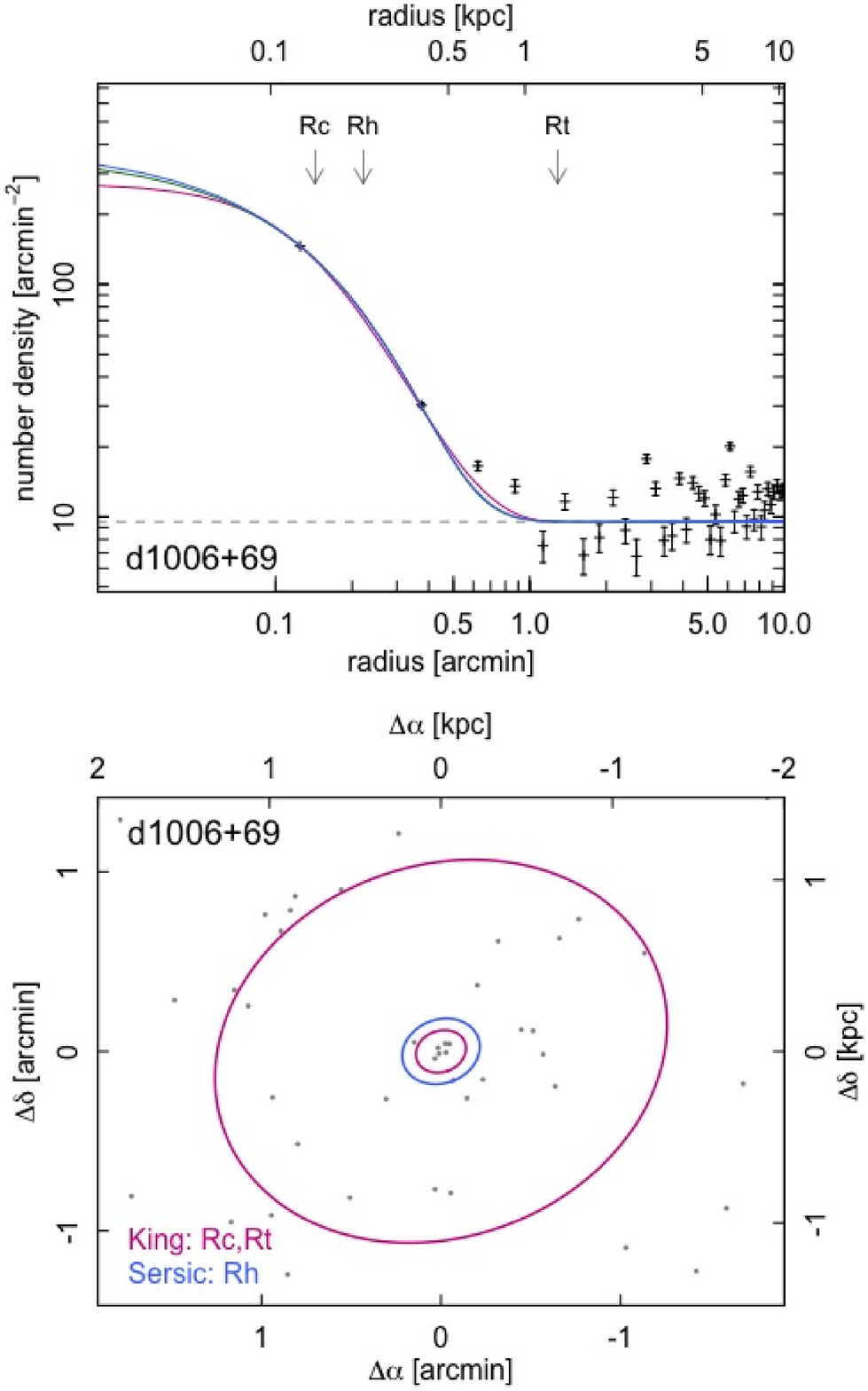}
 \caption{Same as Figure \ref{fig: map and profile 1}, but for the dwarf galaxy candidate d1006+69.}
 \label{fig: map and profile UKO01}
\end{center}
\end{figure}

\begin{figure}
\begin{center}
 \includegraphics[width=170pt]{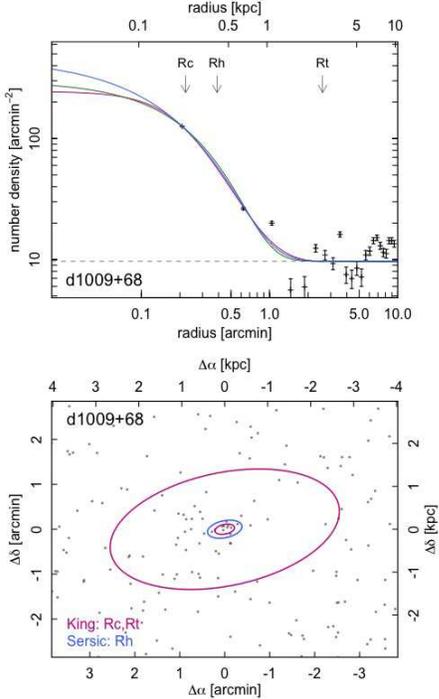}
 \caption{Same as Figure \ref{fig: map and profile 1}, but for the dwarf galaxy candidate d1009+68.}
 \label{fig: map and profile UKO02}
\end{center}
\end{figure}

Figure \ref{fig: uko01 cmd} and \ref{fig: uko02 cmd} show the CMDs of galaxies and their nearby reference fields.  The panels show stellar objects lying within the core radius $\rc$ (left), the Sersic half-light radius $\rh$ (middle), and a reference field of radius $\rh$ centerd $5\arcmin$ away from the galaxy center (right). The theoretical PARSEC v1.2 isochrones of 10 Gyr old population with $\gmh=-2.2, -1.75, -1.3, -0.75$
are overlaid.  Blue RGBs can be seen in the left and middle panels of both figures, which are similar to those of the
M81 dwarf satellites d1015+69 and d1005+68, while no RGB sequence is apparent in the reference field.  The photometric and structural properties are estimated in exactly the same manner as those in Section \ref{subsec: old dwarf structure} and reported in Table \ref{tbl: str UKO}.  Figure \ref{fig: map and profile UKO01} and \ref{fig: map and profile UKO02} show the radial profiles and the spatial distributions of RGB stars; the
slight apparent excess of stars at $> \rh$ is unlikely to be significant.  The average metallicities $\langle\gmh_{w}\rangle = -1.83 \pm 0.28$ and $\langle\gmh_{w}\rangle = -1.47 \pm 0.29$ are derived from the RGB colour in the same manner as that done in Section \ref{subsec: old dwarf pop}.  If confirmed as dwarf galaxies,  they will be one of the faintest satellites known in the M81 Group. 

\begin{deluxetable}{ccc}
\tabletypesize{\scriptsize}
\tablecolumns{3}
\tablewidth{0pt}
\tablecaption{The properties of new systems\label{tbl: str UKO}}
\tablehead{
\colhead{Parameter} & \colhead{d1006+69} & \colhead{d1009+68} }
\startdata
R.A.(J2000)    & $10^{h}06^{m}55^{s}.5$ & $10^{h}09^{m}14^{s}.3$ \\
Dec. (J2000)   & $+69\arcdeg54\arcmin16\arcsec.6$ & $+68\arcdeg45\arcmin24\arcsec.6$ \\
Position Angle [deg] & $110.0$ & $101.7$ \\
$\epsilon$     & 0.20 & 0.51\\
$\mm$          & $28.18\pm0.10$  & $27.86\pm0.29$\\
Distance [Mpc]      & $4.33\pm0.20$ Mpc & $3.7\pm0.5$ Mpc\\
$\rc$[arcmin]          & $0.14\pm 0.07$ & $0.22\pm 0.11$  \\
$\rt$[arcmin]          & $1.3\pm0.9$ & $2.6\pm2.0$\\
$\re$[arcmin]          & $0.13\pm0.01$ & $0.23\pm0.03$ \\
n              & $1.04\pm0.96$ & $1.28\pm1.47$ \\
$\rh$[arcmin]          & $0.22\pm0.03$ & $0.39\pm0.09$\\
$\mv$          & $-8.91\pm0.40$ & $-8.73\pm0.45$\\
$\vi$       & $1.24\pm0.26$ & $1.18\pm0.30$\\
$\langle\gmh_{w}\rangle$  & $-1.83\pm0.28$ & $-1.43 \pm 0.28$\\
$\sigma_{\gmh}$ & $0.16$  & $0.49$ \\
\enddata
\end{deluxetable}

\section{Young stellar systems} \label{sec: young dwarfs}

\begin{figure*}
\begin{center}
 \includegraphics[width=450pt]{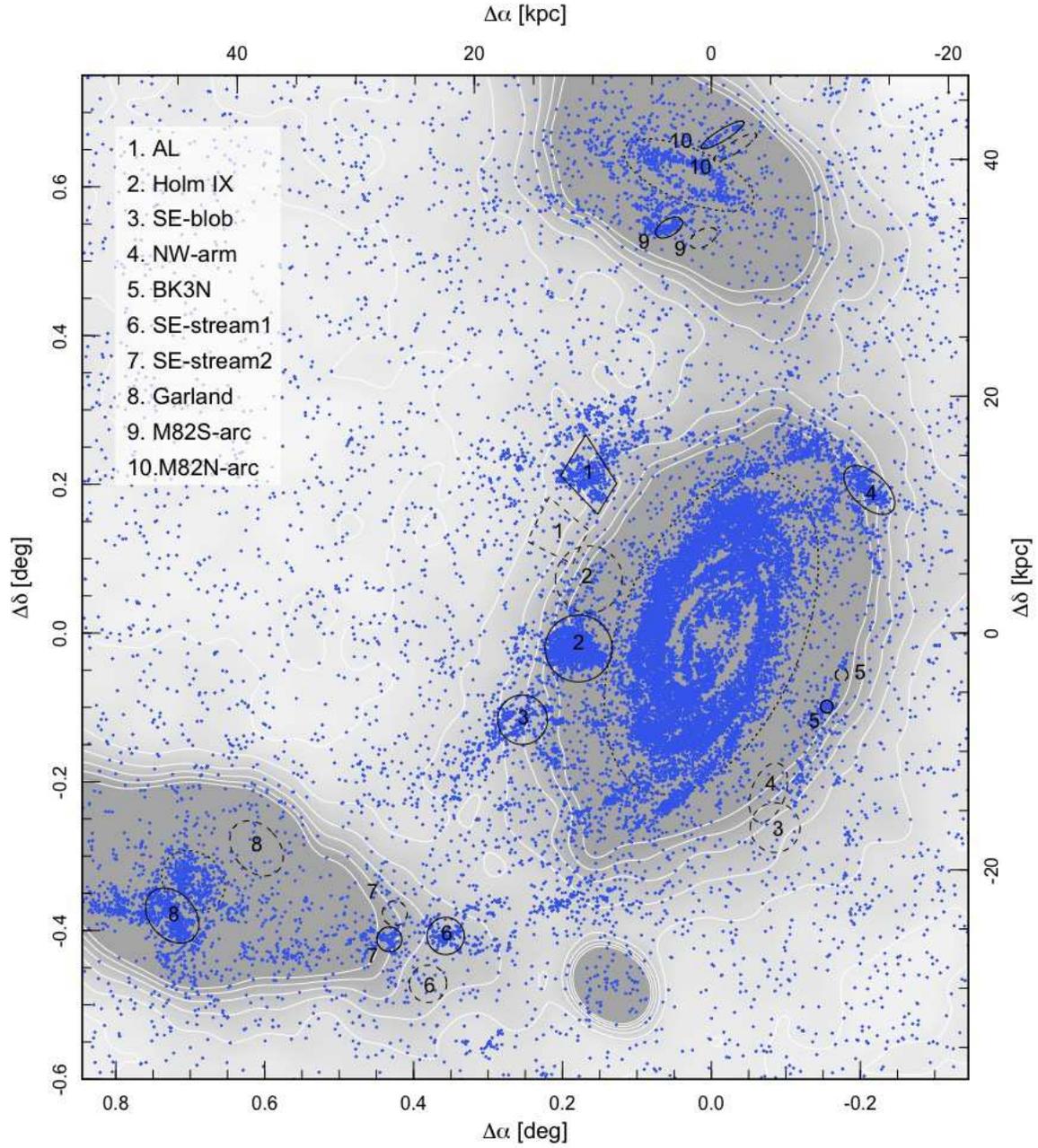}
 \vspace{5pt}
 \caption{Spatial map of resolved stars in the core of the M81 Group. The solid and dashed regions indicate the target and comparison fields of the young stellar systems studied in Section \ref{sec: young dwarfs}, respectively. The blue points represent the young MS, cHeB, and RSG stars and they are plotted on top of a contour map of the RGB stellar density created with the kernel bandwidth of 1.2\arcmin. The dotted lines show the $\rr$ radii of M81, M82, and NGC3077. }
 \label{fig: map of young dwarfs}
\end{center}
\end{figure*}

Tidal interactions between M81, M82, and NGC3077 have stripped gas from the main bodies of
these systems and a number of young stellar clumps and streams are known to coincide with the 
densest concentrations of neutral hydrogen gas \citep{1985MNRAS.217..731K, 2004IAUS..217...90D, 2005ApJ...630L.133S, 2008PASP..120.1145D, 2009MNRAS.399..737M, 2013AJ....146..126C, 2015ApJ...809L...1O}.  In this section, we focus on the following ten areas as shown in Figure \ref{fig: map of young dwarfs}; (1) Arp's Loop (AL), (2) Holmberg IX (Holm IX), (3) an over-density at the southeast of Holm IX (SE-blob), (4) an arm at the northwest side of the M81 outer disk (NW-arm; identified as the ``M81 West" by \citet{2005ApJ...630L.133S}), (5) BK3N, (6) and (7) two denser regions on the stellar stream (SE-stream) at the southeast of M81 and in between M81 and NGC3077 (SE-stream1 and SE-stream2), (8) the Garland, (9) and (10) arcs at the south and the north of M82 (M82S-arc, M82N-arc).   
Most of them are previously-identified prominent star forming debris features around M81, M82 and NGC3077 \citep{1965Sci...148..363A, 1969ArA.....5..305H, 2005ApJ...630L.133S, 1982AN....303..189B, 2015ApJ...809L...1O, 1974A&A....35..463B, 2005ApJ...630L.133S}. 
The blue points represent the young main-sequence (MS), core Helium-burning (cHeB), and red super giant (RSG) stars and the underlaid contours show the distribution of both red and blue RGB stars. The colour-magnitude selection criteria for these populations are indicated in Figure \ref{fig: icmds of young dwarfs}. A
zoom-in view of the stellar distribution in each target is shown in Figure \ref{fig: maps of young dwarfs}, and the coordinates of their central positions are listed in Table \ref{tbl: young systems}.  We also select comparison fields that are located at the same elliptical radius from the nearest large galaxy (M81, M82, or NGC3077) as the target fields and which are as free as possible of stellar enhancements. These comparison fields are used to estimate and correct for the smooth stellar envelope of the nearest large galaxy at the target areas.   While these ten regions represent the most prominent young associations in the core of the M81 Group, several other concentrations of young stars are visible in Figure \ref{fig: map of young dwarfs} but we postpone discussion of these to a later paper.  

\begin{deluxetable}{clcc}
\tabletypesize{\scriptsize}
\tablewidth{0pt}
\tablecaption{The coordinates of young systems \label{tbl: young systems}}
\tablehead{
\colhead{no.} & \colhead{Name} & \colhead{$\alpha$ (J2000)} & \colhead{$\delta$ (J2000)} }
\startdata  
1 & Arp's Loop       & $09^{h}57^{m}22^{s}.9$ & $+69\arcdeg16\arcmin36\arcsec$  \\
2 & Holmberg IX     & $09^{h}57^{m}33^{s}.3$ & $+69\arcdeg02\arcmin45\arcsec$  \\
3 & SE-blob   & $09^{h}58^{m}24^{s}.0$ & $+68\arcdeg57\arcmin00\arcsec$  \\
4 & NW-arm    & $09^{h}53^{m}11^{s}.0$ & $+69\arcdeg15\arcmin32\arcsec$  \\
5 & BK3N     & $09^{h}53^{m}49^{s}.4$ & $+68\arcdeg58\arcmin04\arcsec$  \\
6 & SE-stream1   & $09^{h}59^{m}33^{s}.3$ & $+68\arcdeg39\arcmin33\arcsec$  \\
7 & SE-stream2   & $10^{h}00^{m}24^{s}.0$ & $+68\arcdeg39\arcmin17\arcsec$  \\
8 & Garland     & $10^{h}03^{m}40^{s}.1$ & $+68\arcdeg41\arcmin12\arcsec$  \\
9 & M82S-arc    & $09^{h}56^{m}11^{s}.8$ & $+69\arcdeg36\arcmin42\arcsec$  \\
10 & M82N-arc    & $09^{h}55^{m}23^{s}.4$ & $+69\arcdeg44\arcmin11\arcsec$  \\
\enddata
\end{deluxetable}

\subsection{CMDs}

\begin{figure*}
\begin{center}
 \includegraphics[width=440pt]{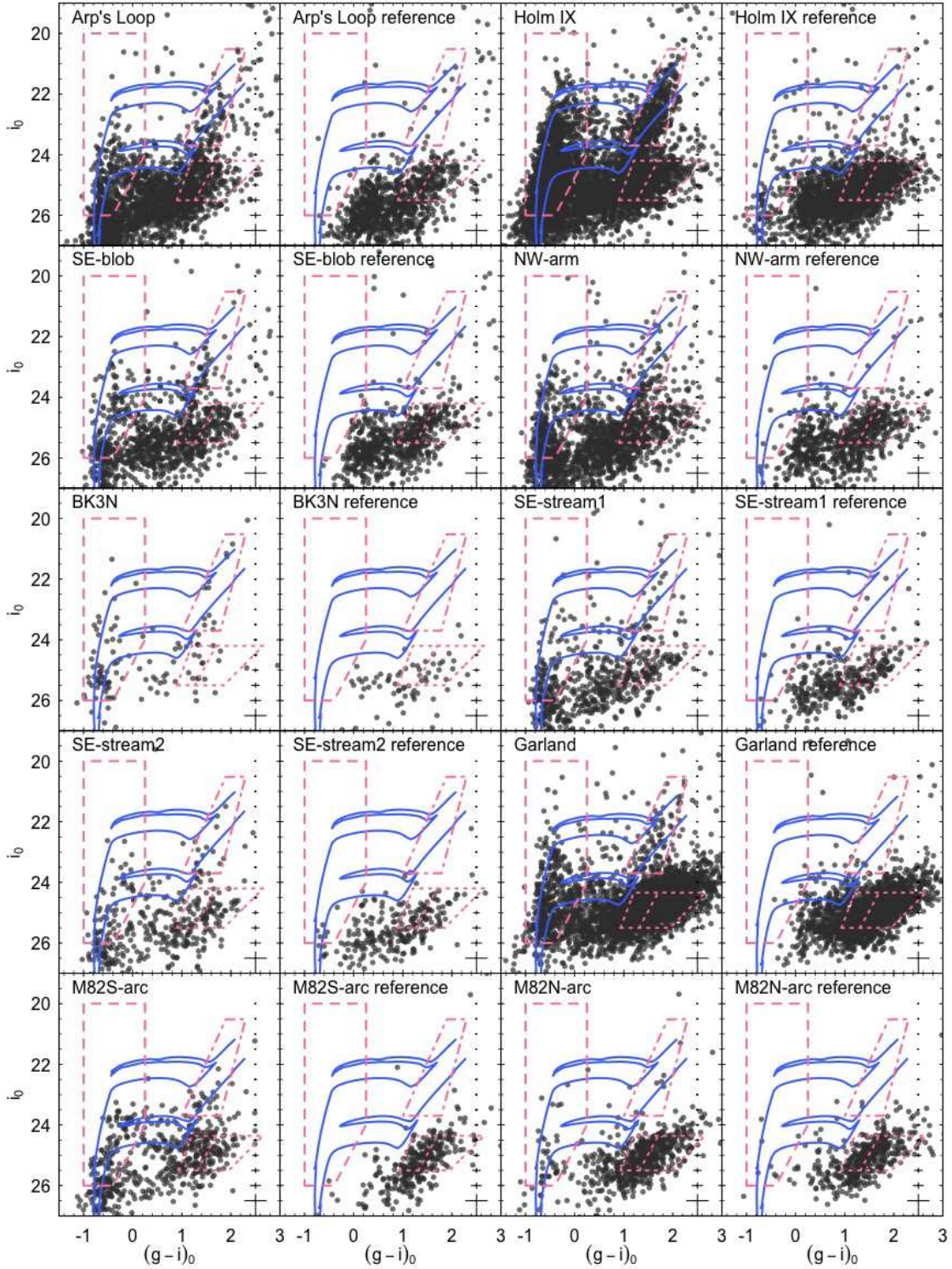}
 \vspace{5pt}
 \caption{The de-reddened $\oi$ versus $\ogi$ CMDs of stars in the young associations and their reference fields. The positions of the target and reference fields are presented in Figure \ref{fig: map of young dwarfs}.  Theoretical isochrones of 32 Myr and 100 Myr with $\gmh=-0.75$ are overlaid as blue solid lines. The error bars indicate the photometric error at $\ogi=0$ at each galaxy and its reference field.  Stars within dotted, dashed-dotted and dashed lines are used to map the spatial distributions of red and blue RGB, RSG, and MS/cHeB stars, respectively in Figures \ref{fig: maps of young dwarfs}, \ref{fig: maps of pops in youngdwarfs1} and \ref{fig: maps of pops in youngdwarfs2}.}
 \label{fig: icmds of young dwarfs}
\end{center}
\end{figure*}

\begin{figure}
\begin{center}
 \includegraphics[width=260pt]{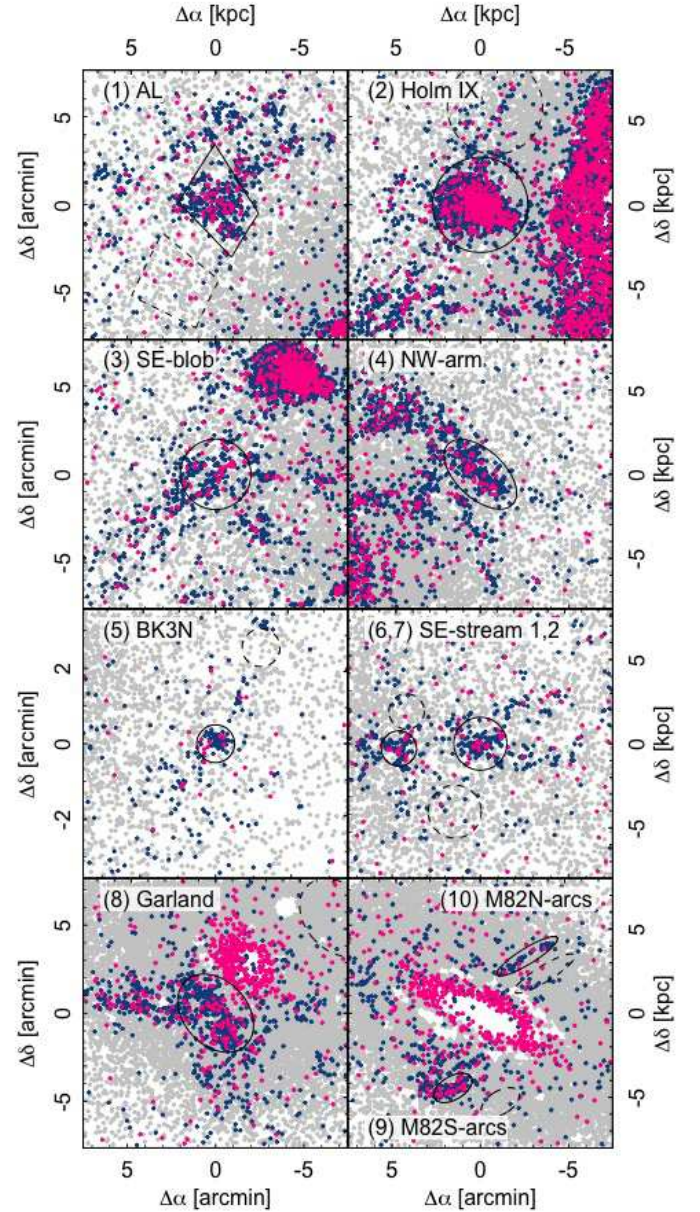}
 \caption{The spatial distribution of stars in the young stellar systems. The grey, magenta and dark blue points represent the RGB, RSG, and MS/cHeB stars selected within the dotted, dashed-dotted and dashed lines in Figure \ref{fig: icmds of young dwarfs}, respectively.  The solid and dashed lines indicate the target and reference fields of young stellar systems, respectively.}
 \label{fig: maps of young dwarfs}
\end{center}
\end{figure}

\begin{figure*}
\begin{center}
 \includegraphics[width=440pt]{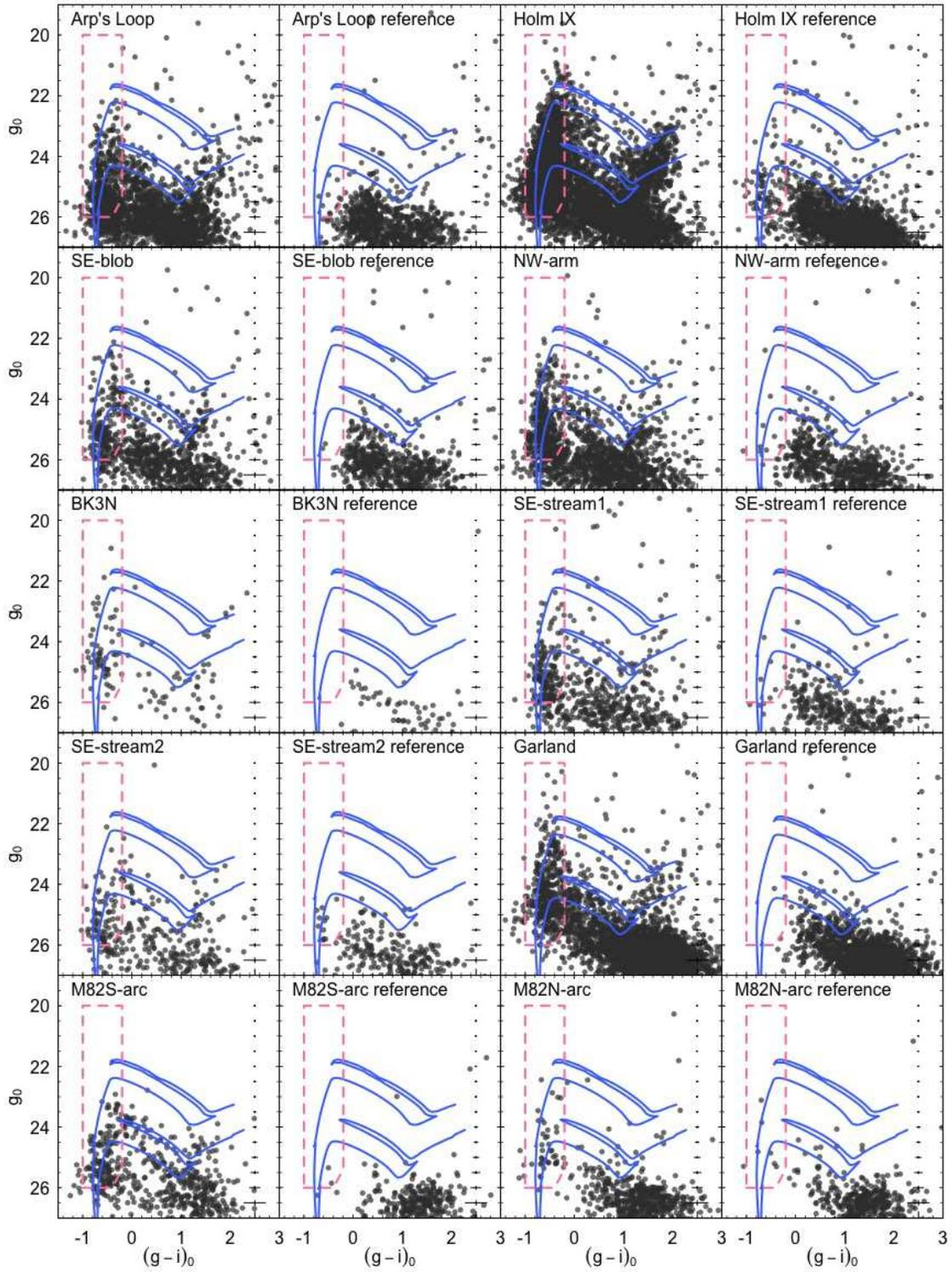}
 \vspace{5pt}
 \caption{Same as Figure \ref{fig: icmds of young dwarfs}, but the $\og$ versus $\ogi$ CMDs. The dashed line shows the selection box of MS stars used to derive the luminosity functions in Figure \ref{fig: LF of young dwarfs}. }
 \label{fig: gcmds of young dwarfs}
\end{center}
\end{figure*}

\begin{figure}
\begin{center}
 \includegraphics[width=220pt]{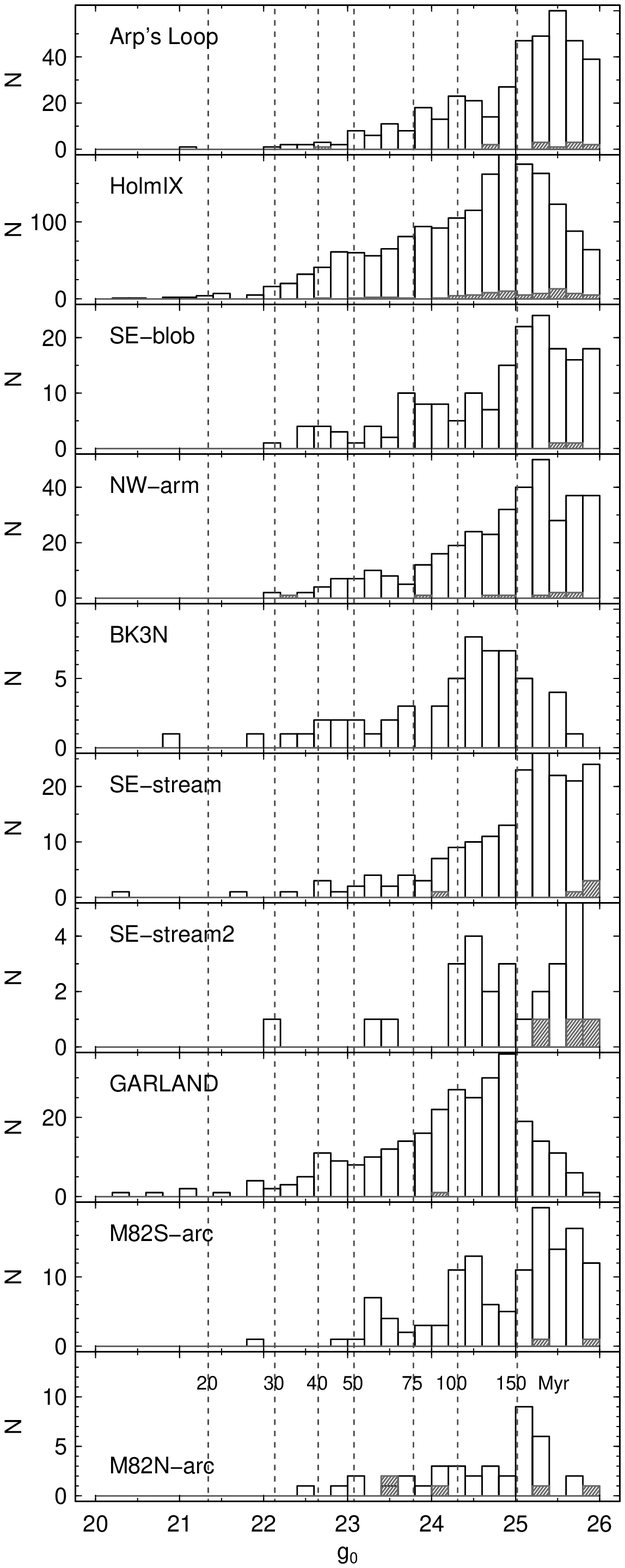}
 \caption{The $g$-band luminosity functions of young stellar systems within the selection boxes of Figure \ref{fig: gcmds of young dwarfs}.  The grey shaded histograms show the luminosity function of  contaminants as estimated from the reference fields. The vertical dashed lines indicate the $g$-band magnitudes of the brightest MS/cHeB stars in isochrones of 20 to 150 Myr old with $\gmh=-0.75$,  adjusted to the distance of M81. }
 \label{fig: LF of young dwarfs}
\end{center}
\end{figure}

Figure \ref{fig: icmds of young dwarfs} shows the CMDs of young stellar systems listed above.  In the CMDs for AL and Holm IX,  shown in the upper panels of Figures \ref{fig: icmds of young dwarfs}, the most prominent features are vertical distributions at $\ogi\sim-0.5$ composed of MS and cHeB stars, and another vertical distribution at $\ogi\sim1.8$ composed of RSG stars.  
The colour and magnitude distributions of RGBs in AL and Holm IX are similar to those in the reference CMDs.   The red RGBs (92 stars) in AL CMD are comparable to those in its reference CMD (107 stars), while those in Holm IX are less than those in its reference due to crowding.  This implies that the old populations present in these fields are simply due to the M81 halo (see also \citet{2008AJ....135..548D, 2008ApJ...676L.113S,2009AJ....138.1469B}); this is discussed further in Sec \ref{subsec: old stars in young systems}. 
Similar CMD features can be seen in the SE-blob, NW-arm, BK3N, SE-streams, and the Garland. On 
the SE-stream, the CMDs of the two regions are very similar to each other.  In the stellar arcs near 
 M82, the brightest young stars are fainter than those seen in other systems, suggesting older ages. 

Overall, these young stellar systems show clumpy spatial distributions and appear to be embedded in 
much smoother distributions of old RGB stars (see Figure \ref{fig: maps of young dwarfs}).  AL appears to have a tidal structure extending to the northwest side, while Holm IX shows the relatively round shape.  The center of the RSG distribution in Holm IX appears offset from that of the MS/cHeB stars, while other systems do not show such a difference.  Since the CMD of HoIm IX in Figure \ref{fig: icmds of young dwarfs} shows more bright MS/cHeB stars in comparison with other systems, the spatial difference of RSG and MS/cHeB in HoIm IX implies the different recent star formation history in this area from others.  The SE-blob is elongated towards the southeast while the  NW-arm, which lies at an elliptical radius of 22 kpc from the M81 center, appears to truncate abruptly.  BK3N, the smallest of the young stellar systems studied here, is located at the same elliptical radius as NW-arm and may be a further fragment of that feature.  The SE-stream, lying between M81 and NGC3077, shows regions of local enhancement.  The Garland has a very unique shape, being highly elongated towards the south and the east.  Interestingly,  there is no obvious counterpart to this in RGB distribution.  The stellar arcs of M82 lie perpendicular to the disk. Although the width of the M82S-arc is much broader the M82N-arc, their positions appear to be axially symmetric.  

\subsection{The Luminosity Functions}

To further investigate the stellar populations in these systems, we also plot the $\og$ versus $\ogi$ CMDs in Figure \ref{fig: gcmds of young dwarfs}.  Compared with the $\oi$ versus $\ogi$ CMDs, the young MS/cHeB stars in Figure \ref{fig: gcmds of young dwarfs} have the luminosity peak at around $\ogi=-0.4$ then their evolutionary track moves redder and fainter. The $g$-band luminosity of the brightest star is highly dependent on age but is insensitive to metallicity.  It varies by about 0.25 magnitude between the 
$\gmh=-1.2$ and $-0.2$ isochrones at constant age, while it varies about 4.5 magnitudes between ages of 0 and 200 Myr isochrones at constant metallicity \citep{2012MNRAS.427..127B}.  Therefore, we investigate the $g$-band LF of upper MS and cHeB stars in order to set a constraint on the duration
of the star formation period. 

Figure \ref{fig: LF of young dwarfs} shows the LF of young stars that lie in the selection box show 
in Figure \ref{fig: gcmds of young dwarfs}.  The vertical dashed lines show the g-band magnitudes of the brightest MS/cHeB stars along isochrones of age 20 to 150 Myr with $\gmh=-0.75$, assuming populations
at the distance of M81. While the Garland and the M82S/N-arcs are likely to be more closely
associated with NGC3077 and M82,  the distance moduli to these systems differ by only
 -0.13 mag, and 0.05 mag from that of M81 \citep{2009ApJS..183...67D}. Given that this is smaller than our  binsize of 0.2 mag, this is unlikely to cause any significant effect.   Photometric incompleteness becomes
 important at roughly $\og>25$, which is manifest in the LF downturns seen in Holm IX, the Garland and BK3N.  We do not correct for the incompleteness in Figure \ref{fig: LF of young dwarfs} since we are solely
 interested in the brighter part of the LF.   

AL, the SE-blob, NW-arm, and SE-stream1 all have similar LFs which decrease gradually with increasing luminosity then flatten out at $\og\sim23$, with few or no stars brighter than $\og=22$.  This suggests that the youngest populations in these systems are $\sim30$ Myr old and that they have all experienced similar recent SFHs.  In the young metal-poor population, the upper MS stars ($\mg<-2.3$, corresponding to $\og<25$ at the M81 distance) and the brightest MS/cHeB stars have similar mass; $\rm{M}=8.1\msun$ to $8.8\msun$ in the 30 Myr old population and $\rm{M}=5.5\msun$ in the 75 Myr old population, which makes a flat LF for a single stellar population at this magnitude range.  Therefore, the moderate slopes of these LFs indicate continuous star formation in these young objects in the period of 50 - 150 Myr ago.  BK3N also has similar LF at $\og<25$.  Considering the locations of these systems as well as the lack of any counterparts in the old population (see Section \ref{subsec: old stars in young systems}), they are likely to have formed near simultaneously from gaseous material stripped from M81.  

The LF of Holm IX decreases gradually from $\og=25$ to $\og\sim22$ where it remains flat up to $\og\sim21$. The presence of brighter stars in this region suggests that it has sustained star formation for a longer period than the other systems.  The Garland has a similar LF to Holm IX, but it has a small bump at $\og\sim22.7$.  Stars in the M82S-arc and M82N-arc features populate the LF to $\og\sim22.5$, implying that they stopped forming stars earlier than the other systems.

\section{Discussion} \label{sec: discussion}

\subsection{Tidal effects on the early-type galaxies}
\label{subsec: tidal effect}

Compared to other nearby galaxy groups, the M81 Group is unique because of the strong gravitational interactions that are ongoing amongst M81, M82 and NGC3077.  While tidal stripping has displaced
gas from the main bodies of these systems and triggered new star formation within it,  we have found little  evidence for tidal stripping 
in the old dwarf galaxy population (for e.g., signatures of extra-tidal stars in their radial profiles or distorted isophotes). 

\citet{2009ApJ...698..222P} investigated the effect of tidal stripping on the structural properties of a dSph galaxy.  They showed that the radial surface brightness profile deviates substantially from a King profile as a result of the tidal interaction. Once the dSph equilibrates again, the outer density profile settles into a power-law varying as $\rm{R}^{-4}$.  An excess of stars with respect to a King profile has been found
in the outer parts of some Milky Way dSphs, such as Fornax and Ursa Minor,  and cited as evidence for tidally stripped populations \citep[e.g.][]{2003AJ....125.1352P, 2005AJ....129.1443C}.  Marginal overdensities can be seen at around $\rt$ in the radial profiles of d1005+68 and d1015+69 shown in Figure \ref{fig: map and profile 2}, but it is arguable whether they are real features or simply due to the poor statistics.   Overall, the old dwarf satellite population seems  to have been unaffected by the strong tidal
interactions which are shaping the properties of the most massive galaxies.

\begin{figure}
\begin{center}
 \includegraphics[width=260pt]{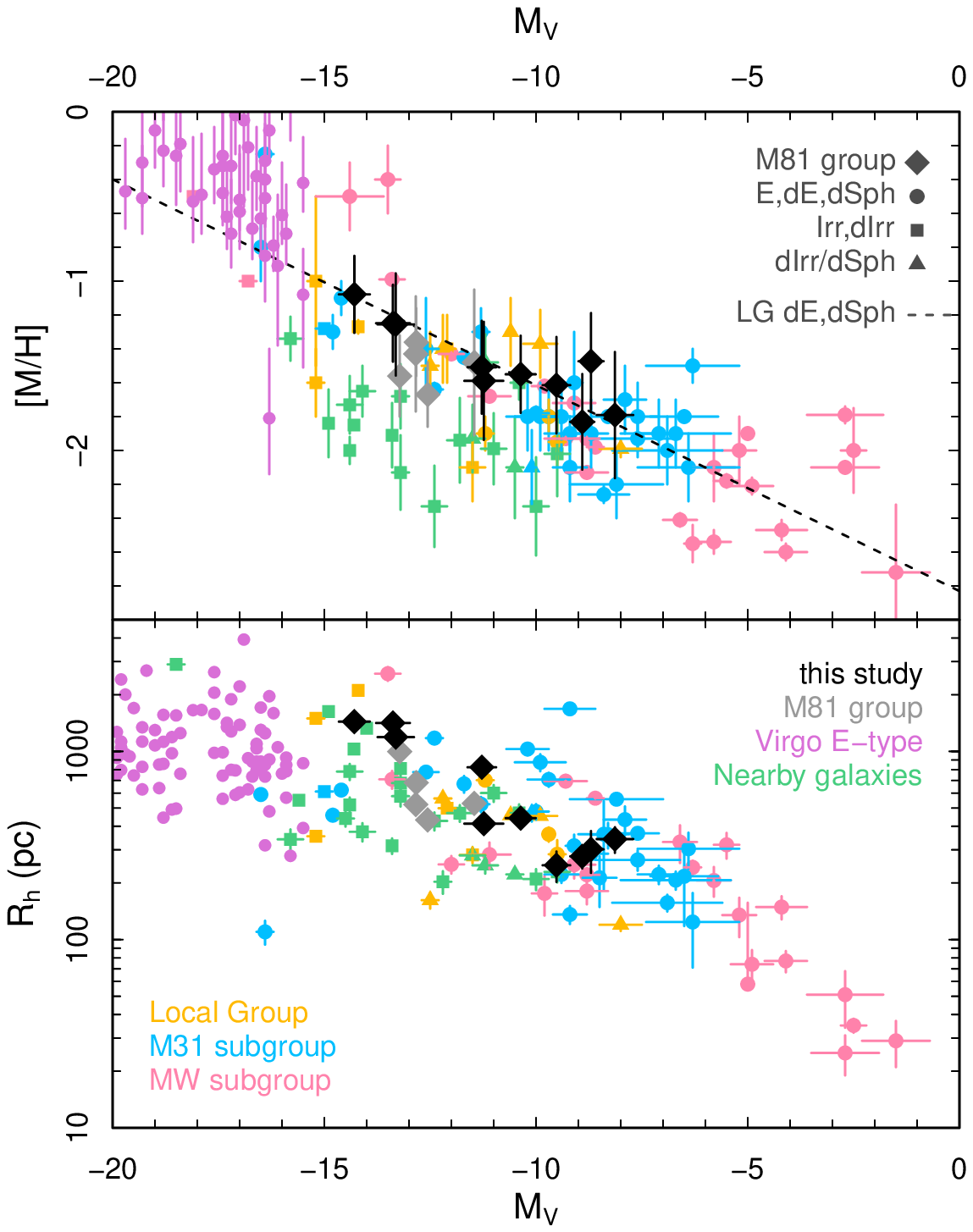}
 \caption{The luminosity-metallicity and the size-luminosity relations of nearby galaxies. The M81 Group galaxies studied here are shown in black diamonds.  Other M81 dwarf galaxies studied by \citet{2010A&A...521A..43L}, nearby galaxies from \citet{2012AJ....144....4M}, and early-type galaxies in  the Virgo cluster from \citet{2011MNRAS.414.3699M} and \citet{2017arXiv170900406S} are also plotted. {\it Upper:} Absolute V-band magnitude vs. metallicity of galaxies.  For the galaxies having previous metallicity and $\rh$ estimation, we use the values estimated in this study.  The dashed line is fitted to dEs and dSphs of the Local Group.  {\it Lower:} Absolute V-band magnitude vs. the half light radius of galaxies. }
\label{fig: dwarfs Mv metal Rh}
\end{center}
\end{figure}

\subsection{Comparison of M81 dwarf galaxies and other nearby dwarf galaxies}
\label{subsec: dsph comparison}

The photometric properties of the dwarf galaxies derived in this study are similar to those of early-type dwarf galaxies in the Local Group.  Figure \ref{fig: dwarfs Mv metal Rh} shows the comparison of the total magnitudes, mean metallicities, and  sizes of dwarf galaxies in the nearby universe.  The M81 Group galaxies studied in Section \ref{sec: old dwarfs} are shown as black diamonds.  Other M81 Group galaxies, DDO44, DDO71, DDO78, F6D1, F12D1 from \citet{2010A&A...521A..43L}, are plotted as grey diamonds.  The filled circles, boxes, and triangles indicate morphological types of dwarf elliptical (dE) or spheroidal (dSph), irregular (Irr) or dwarf irregular (dIrr), and transitional type, respectively, taken from \citet{2012AJ....144....4M}. 

We fit the luminosity-metallicity relation of the dE and dSph galaxies in the Local Group with a linear least-square fit,  shown as the dashed line.  Magenta, cyan, green and yellow colours of the symbols represent Galactic satellites, M31 satellites, dwarf galaxies in the Local Group and in the nearby field ($<3$ Mpc), respectively.  The radii of early-type galaxies in the Virgo cluster taken from \citet{2011MNRAS.414.3699M} and the metallicities of nuclei belonging to these galaxies, taken from \citet{2017arXiv170900406S}, are also presented as purple circles.  Note that the different methods adopted for these metallicity and luminosity measurements will contribute the scatter in Figure \ref{fig: dwarfs Mv metal Rh}.

\begin{figure*}
\begin{center}
 \includegraphics[width=420pt]{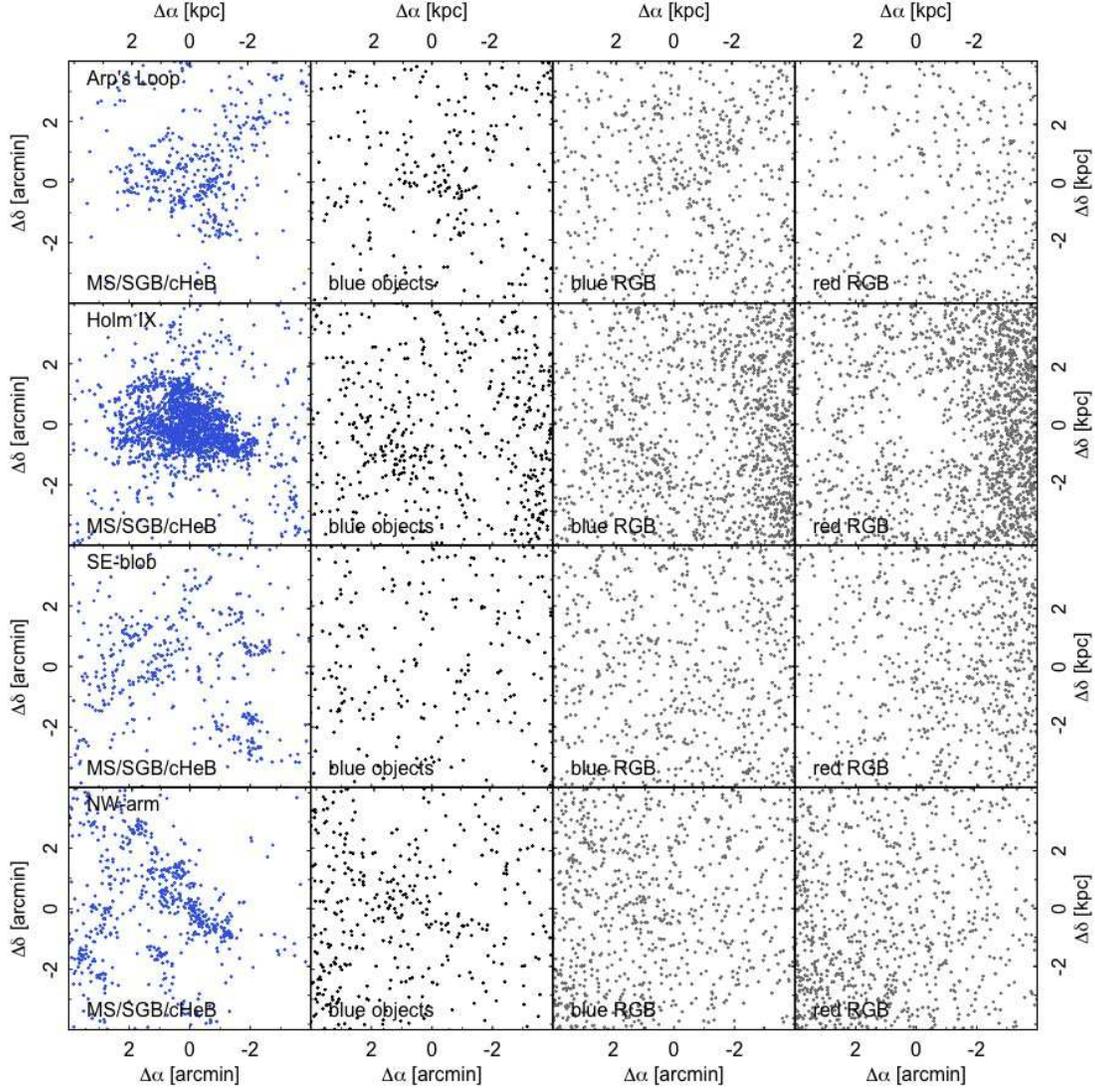}
  \vspace{5pt}

 \caption{The spatial distribution of MS/cHeB stars, stars located between $\ogi=0.75$ and the boundary of the blue RGB selection box, and blue and red RGB stars in AL, Holm IX, SE-blob, and NW-arm. }
 \label{fig: maps of pops in youngdwarfs1}
\end{center}
\end{figure*}

In the luminosity-metallicity plot shown in the upper panel of Figure \ref{fig: dwarfs Mv metal Rh},  M81 Group galaxies can
be seen to obey the same relation as that of dE and dSph galaxies in the Local Group, shown as the dashed line ($\gmh=-0.12\times\mv-2.83$). They appear to have somewhat higher metallicities than nearby isolated dIrr galaxies (green symbols) of  the same total magnitude.  The nuclei of luminous early-type galaxies in the Virgo cluster have higher mean metallicities than the luminosity-metallicity relation of fainter early-type galaxies, conceivably due to a selection effect on the calculation of the mean metallicities through the spatial extent of galaxies.  \citet{2017arXiv170900406S} estimated the mean metallicities from Lick spectral indices of nuclei, where more metal rich stars are likely to be concentrated.  \citet{2010A&A...521A..43L} and \citet{2015A&A...581A..84T} pointed out that IKN has a higher mean metallicity for its luminosity compared to other dE and dSph galaxies in the M81 Group and hence
that IKN may have formed as a tidal dwarf galaxy.  However, the significantly brighter $\mv$ calculated for IKN in this study 
places it on the same metallicity-luminosity relation as for other early-type galaxies of the M81 Group.  This updated luminosity for 
IKN also means that it has a less extreme globular cluster specific frequency \citep{2014A&A...565A..98L}.   

The lower panel of Figure \ref{fig: dwarfs Mv metal Rh} shows the size-luminosity relationship of the dwarf galaxy sample.  The relationships defined by the M31 and MW dSph galaxies are statistically consistent with each other \citep{2011ApJ...743..179B, 2012ApJ...752...45T}.  Our measurements for the M81 Group dwarf galaxies indicate that they also obey the same trend.

\subsection{Are there old stars associated with the young systems?}
\label{subsec: old stars in young systems}

\begin{figure*}
\begin{center}
 \includegraphics[width=420pt]{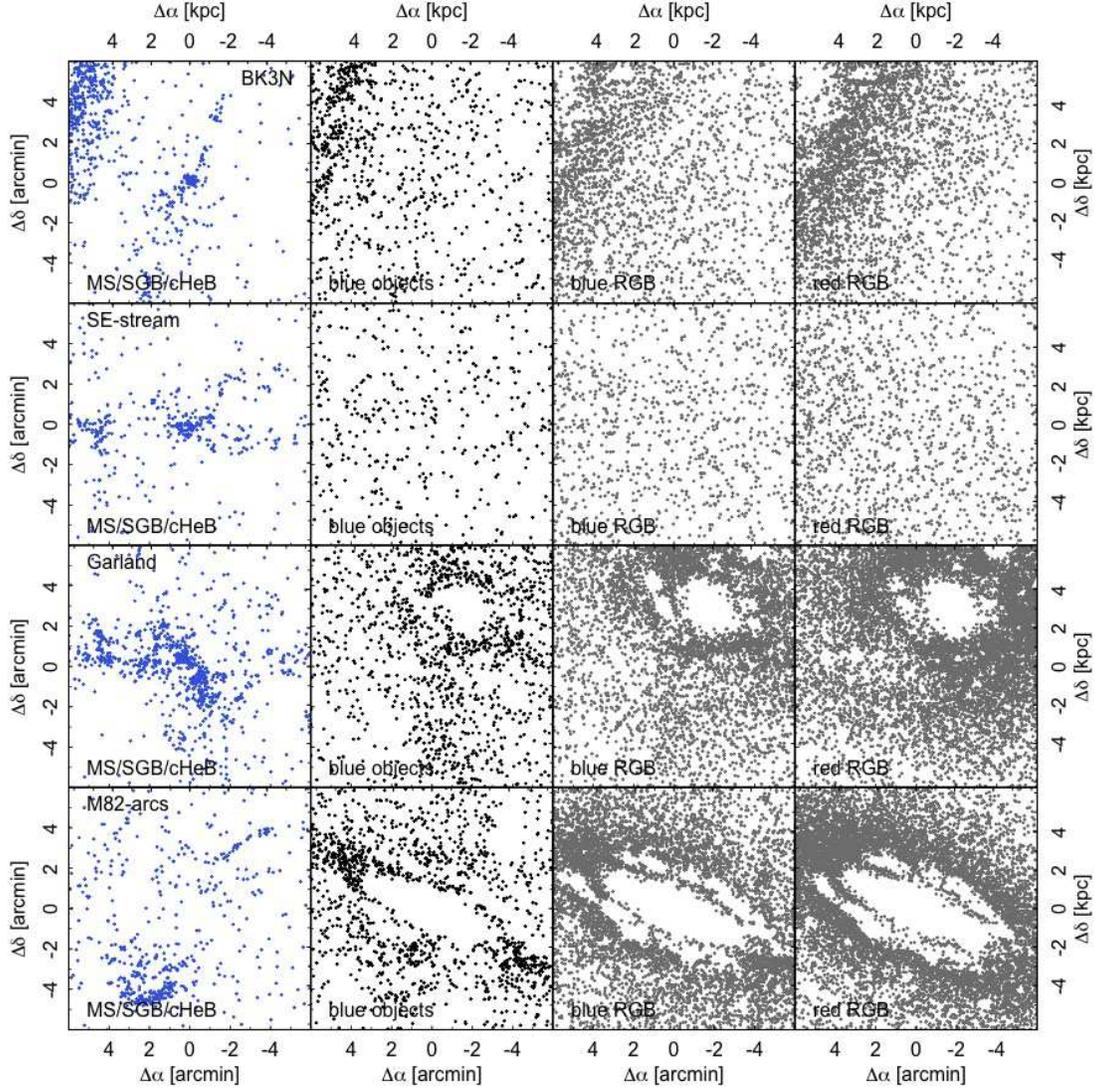}
  \vspace{5pt}
 \caption{Same as Figure \ref{fig: maps of pops in youngdwarfs1}, but for BK3N, SE-stream, the Garland, and M82S/N-arcs.}
 \label{fig: maps of pops in youngdwarfs2}
\end{center}
\end{figure*}

A long-standing question is whether the young stellar systems in the M81 Group 
are purely young stellar associations that have formed in gaseous tidal debris, or 
whether they are gravitationally bound early-type dwarf galaxies that simply have elevated
star formation rates at the present epoch \citep[e.g.][]{1987A&A...188....5H, 1993ApJ...402L..45H}.  Deep HST images and 
chemical abundance measurements support the former idea for Holm IX, AL, BK3N, and the Garland \citep{2002A&A...396..473M, 2008AJ....135..548D, 2008ApJ...676L.113S, 2008ApJ...689..160W, 2009ApJ...705..723C}.  Specifically, \citet{2008ApJ...689..160W} did not identify RGB overdensities at the positions of Holm IX and the Garland in deep HST data, even although such overdensities were associated with the other dwarf galaxies in their sample. \citet{2009AJ....138.1469B} investigated this same question using
deep wide-field photometry taken with Subaru's Suprime-Cam, and also concluded there was no distinct population of RGB stars associated with AL.  In this section, we revisit this question for all of the prominent young stellar systems in the core of
the M81 Group using our deep HSC photometry.  Although our data suffers significant incompleteness due to crowding
in the innermost regions of some systems,  it still enables a useful exploration of the spatial distributions of young and
old stars in the vicinity of the objects discussed in Section \ref{sec: young dwarfs}. 

Figure \ref{fig: maps of pops in youngdwarfs1} and Figure \ref{fig: maps of pops in youngdwarfs2} show the spatial distributions of young MS/cHeB stars, blue and red RGB stars, as well as stars located between $\ogi=0.75$ and the boundary of the blue RGB criterion in Figure \ref{fig: icmds of young dwarfs} (named ``blue objects" in the figures).  The MS/cHeB stars and blue and red RGB stars are selected using the dashed and dotted polygons shown in Figure \ref{fig: icmds of young dwarfs}.  The apparent absence of RGB stars at the center of Holm IX (Figure \ref{fig: maps of pops in youngdwarfs1}) as well as at the centers of M82 and NGC3077 (see Figure \ref{fig: maps of pops in youngdwarfs2}) is due to the severe incompleteness due to crowding. 

There are no obvious counterparts of the young systems in the red RGB distributions in Figure \ref{fig: maps of pops in youngdwarfs1} and Figure \ref{fig: maps of pops in youngdwarfs2}.  In the blue RGB maps, there is marginal evidence for diffuse enhancements at 
the positions of AL and the Garland.  However, it is unlikely that these are very metal-poor ($\gmh\sim-2$) and old ($> 10$ Gyr) stars associated with young objects; they are more likely to be ``young" blue loop stars of ages 100 to 400 Myr old that fall 
within the blue RGB selection box of Figure \ref{fig: icmds of young dwarfs}.  Indeed, similar structures can be
seen in the blue object maps of these systems, which will also capture blue loop stars as well as unresolved background galaxies. 
 Overdensities of blue objects, likely due to blue loop stars,  can be also seen in Holm IX, the NW-arm, and the M82S-arc.   

In summary, our spatial distributions of stars in Figure \ref{fig: maps of pops in youngdwarfs1} and Figure \ref{fig: maps of pops in youngdwarfs2} reveal no evidence for old populations associated with the young systems, down to the limit of our photometry.  This
adds further support to the idea that these are genuinely new stellar systems which have formed in gaseous tidal debris, as
previously suggested for Holm IX, the Garland and AL, the most prominent examples of such features known  \citet{2008ApJ...689..160W, 2009AJ....138.1469B}.   If these structures have internal kinematics that indicate they are self-gravitating, then
they can be considered as tidal dwarf galaxies that will likely survive for a long time. \citep[e.g.][]{2000AJ....120.1238D, 2001AJ....122.2969H}.  

\section{Summary} \label{sec: summary}
The structural and photometric properties of early-type dwarf galaxies and young stellar systems located in the central region of the M81 Group  are investigated in this article.  Using our deep wide-field survey data from HSC, we are able to conduct the
first homogeneous analysis of this sample based solely on their resolved stellar populations. 

We derive the centroids, ellipticities, radii, total luminosities, and metallicities of eight previously-known early-type dwarf galaxies;  IKN, KDG061, KDG064, BK5N, d0955+70, d1005+68, d1014+68, and d1015+69.  We also conduct the same analysis on d1006+69 and d1009+68, 
new satellite candidates that we present the discoveries of.  With an estimated distance of  $4.3\pm0.2$ Mpc, we find
d1006+69 to be one of the faintest and most metal poor galaxies currently-known in the M81 Group.  The estimated radii of most 
the galaxies in our sample are larger than those derived previously, mostly on the basis of integrated light surface photometry. The radial profiles are well-described by Sersic and King profiles, and show no obvious features of the tidal stripping.  Of particular
note, we find that the the total luminosity of IKN ($\mv=-14.29$) is almost 3 magnitudes brighter than previous estimates ($\mv=-11.51$) based on surface photometry, elevating this system to the brightest dwarf satellite in the M81 Group.  In addition, 
we find that the declination of IKN is also discrepant with that  listed in NED, probably due to the existence of bright foreground stars which have complicated prior analyses of this object.  The metallicity distributions of the galaxies are estimated from the colour of individual RGB stars and we find the shapes of these can be fit with a single Gaussian, similar to those of the luminous Galactic dSphs.  The global properties of the early-type galaxies in the M81 Group follow the same trends in luminosity, metallicity and size as those defined by the dE and dSph galaxies of the Local Group. 

For the young stellar systems that are associated with the tidal HI debris around M81, M82, and NGC3077, we focus on 10 prominent
regions including the well-studied objects  AL and Holm IX.  The CMDs and LFs of AL, the SE-blob, the NW-arm, BK3N and the SE-stream1 indicate continuous star formation between 50 to 150 Myr ago and the existence of populations as young as 30 Myr old.  The LFs of Holm IX and the Garland suggest they have experienced more prolonged periods of SF than the above systems.   We find no
evidence for old RGB stars coincident with the young MS/cHeB stars which define these objects, supporting
the idea that they are genuinely new stellar systems resulting from triggered star formation in gaseous tidal debris.  Measuring the internal kinematics of these objects will be crucial in determining whether they will become long-lived systems or not. 

In future articles, we will use our survey data to examine the stellar halos of the main member galaxies, analyse the
tidal structure around NGC3077 and search for new dwarf galaxy members of the M81 Group. 

\acknowledgments
We are grateful to the entire staff at Subaru Telescope and the HSC team.  We acknowledge the importance of Maunakea within the indigenous Hawaiian community and with all respect say mahalo for the use of this sacred site. 
This paper makes use of software developed for the LSST.  We thank the LSST Project for making their code available as free software at http://dm.lsstcorp.org.  IRAF is distributed by the National Optical Astronomy Observatory, which is operated by the Association of Universities for Research in Astronomy (AURA) under a cooperative agreement with the National Science Foundation.  S. O. acknowledges support in part from MEXT Grant-in-Aid for Research Activity start-up (18H05875) and thanks the Institute for Astronomy at the University of Edinburgh for visitor support during the completion of this work.  We are thankful to M.G. Lee for his continuous interest and fruitful discussions.  
N.A. thanks for the Brain Pool program for financial support, which is funded by the Ministry of Science and ICT through the National Research Foundation of Korea (2018H1D3A2000902). 

\bibliographystyle{apj}
\bibliography{apj-jour,ms}


\end{document}